\documentclass[aps,twocolumn,superscriptaddress]{revtex4}

\usepackage{graphicx}
\usepackage{amssymb}
\usepackage{amsmath}
\usepackage{amsfonts}
\usepackage{color}
\usepackage{hyperref}

\DeclareMathAccent{\ring}{\mathalpha}{operators}{"17}
\providecommand{\st}[1]{_{\text{#1}}}

\providecommand{\ut}[1]{^{\text{#1}}}

\def\onehalf{\frac{1}{2}}

\def\bra{\ensuremath{\langle}}
\def\ket{\ensuremath{\rangle}}

\def\ueq{\ut{eq}}

\def\tr{\mathrm{Tr}}
\def\Imat{\mathbb{I}}
\def\im{\mathrm{i}}

\def\csph{c_s}
\def\cslb{\sigma_s}

\def\kv{\bv{k}}
\def\uv{\bv{u}}

\def\cv{\bv{c}}

\def\rv{\bv{r}}
\def\b0{\bv{0}}

\def\ra{\rightarrow}

\newcommand{\bitem}{\begin{itemize}}
\newcommand{\eitem}{\end{itemize}}
\newcommand{\benum}{\begin{enumerate}}
\newcommand{\eenum}{\end{enumerate}}
\newcommand{\bblock}[1]{\begin{block}{#1}}
\newcommand{\eblock}{\end{block}}
\newcommand{\bmini}[1]{\begin{minipage}{#1}}
\newcommand{\emini}{\end{minipage}}
\newcommand{\btab}[1]{\begin{tabular}{#1}}
\newcommand{\etab}{\end{tabular}}
\newcommand{\btabn}[1]{\begin{tabular}{#1}}
\newcommand{\etabn}{\end{tabular}}
\newcommand{\beq}{\begin{equation}}
\newcommand{\eeq}{\end{equation}}
\newcommand{\beqn}{\begin{equation*}}
\newcommand{\eeqn}{\end{equation*}}
\newcommand{\bmult}{\begin{multline}}
\newcommand{\emult}{\end{multline}}
\newcommand{\bsplit}{\begin{split}}
\newcommand{\esplit}{\end{split}}

\newcommand{\bv}[1]{\mathbf{#1}}

\begin{document}
\thispagestyle{plain}
\title{Thermal fluctuations in the lattice Boltzmann method for non-ideal fluids}
\author{M. Gross}
\email{markus.gross@rub.de}
\affiliation{Interdisciplinary Centre for Advanced Materials Simulation (ICAMS), Ruhr-Universit\"at Bochum, Stiepeler Strasse 129, 44801 Bochum}
\author{R. Adhikari}
\affiliation{The Institute of Mathematical Sciences, CIT Campus, Chennai 600113, India}
\author{M. E. Cates}
\affiliation{SUPA, School of Physics and Astronomy, University of Edinburgh, JCMB Kings Buildings, Mayfield Road, Edinburgh EH9 3JZ, United Kingdom}
\author{F. Varnik}
\affiliation{Interdisciplinary Centre for Advanced Materials Simulation (ICAMS), Ruhr-Universit\"at Bochum, Stiepeler Strasse 129, 44801 Bochum}
\affiliation{Max-Planck Institut f\"ur Eisenforschung, Max-Planck Str.~1, 40237 D\"usseldorf, Germany}

\begin{abstract}
We introduce thermal fluctuations in the lattice Boltzmann method for non-ideal fluids.
A fluctuation-dissipation theorem is derived within the Langevin framework and applied to a specific lattice Boltzmann model that approximates the linearized fluctuating Navier-Stokes equations for fluids based on square-gradient free energy functionals.
The obtained thermal noise is shown to ensure equilibration of all degrees of freedom in a simulation to high accuracy.
Furthermore, we demonstrate that satisfactory results for most practical applications of fluctuating hydrodynamics can already be achieved using thermal noise derived in the long wavelength-limit.
\end{abstract}


\maketitle

\section{Introduction}
Owing to its flexibility and easily parallelizable nature, the lattice Boltzmann (LB) method has by now become an established tool for solving the Navier-Stokes equations for simple as well as complex fluids \cite{succi_book}.
The simulation of systems with phase coexistence is not only important for applications, such as wetting and thin-films \cite{rauscher_wetting_review, bonn_wetting_review}, but also interesting from a theoretical point of view \cite{hohenberg_halperin, bray_phaseordering}. It is well known that the thermal motion of the fluid particles becomes relevant already below the micro-scale, leading---for example---to the Brownian motion of suspended solid particles \cite{fox_uhlenbeck_1970a, hauge_brownian_1973, ladd_1993}.
But also in pure fluid systems, thermal noise plays an important role close to phase transitions \cite{Chaikin_book} or hydrodynamic instabilities \cite{sengers_book, kadau_pnas2007}, and has recently been shown to also have significant effects on nanoscopic free-surface flows \cite{moseler_nanojets_2000, stone_spreading_2005, fetzer_dewetting_2007}. Simulation of such behavior with a deterministic method, such as LB, requires the inclusion of explicit noise sources in the underlying equations.
In this work, we will discuss how thermal noise can be modeled within the LB method for non-ideal fluids.

The history of thermal fluctuations in the Boltzmann equation dates back to Kadomtsev \cite{kadomtsev_1957}, who first applied the Langevin approach to the Boltzmann equation of a dilute gas. It was shown later by Bixon and Zwanzig \cite{bixon_zwanzig_1969}, and independently by Fox and Uhlenbeck \cite{fox_uhlenbeck_1970b}, that this approach in fact leads to the well-known equations of fluctuating hydrodynamics \cite{Landau_FluidMech59} in the limit of large length and time scales. Generalizations of the Boltzmann-Langevin equation to non-ideal gases have been discussed by Klimontovich \cite{klimontovich_nonideal_1973}. It has been shown by Kim and Mazenko \cite{KimMazenko_1991}, that the expressions for the fluctuating stress tensor known for a simple fluid essentially remain valid also for a fluid described by a square-gradient free energy functional.

In the context of the LB method, so far, only the fluctuating ideal gas model has been studied systematically, starting with the work of Ladd \cite{ladd_1994a, ladd_1994b}. There, the Landau-Lifshitz theory of fluctuating hydrodynamics \cite{Landau_FluidMech59} was implemented by adding a fluctuating component to the LB stress modes.
While the model satisfied the fluctuation-dissipation theorem (FDT) at the hydrodynamic level, it was soon realized that it failed to give full equilibration of momentum \cite{laddVerberg_2001}. As first pointed out by Adhikari et\ al.\ \cite{adhikari_fluct_2005}, in order to ensure correct equipartition of fluctuation energy at all length scales, noise must not only be added to the stresses, but to all dissipative modes that exist for a given model. This was confirmed subsequently by D\"unweg et al.\ \cite{duenweg_statmechLB_2007} using a lattice gas analogy.
Notably, Dufty and Ernst \cite{dufty_lblangevin_1993} gave the first general treatment of LB-Langevin models, including a derivation of the appropriate FDT.
We finally mention that there also exist a number of finite-volume schemes for the direct integration of the fluctuating Navier-Stokes equations \cite{bell_methods_2007}.

In the present work, we study thermal fluctuations in a non-ideal fluid within the Langevin framework.
First, the fluctuating hydrodynamic equations for a fluid based on a square-gradient free energy functional and the associated FDT are discussed for a continuum system.
Next, we present a derivation of the FDT appropriate to the non-ideal fluid LB method. Consistency requires that the stochastic LB equation (LBE) leads to the same form of the fluctuating stress tensor that is required by the FDT derived independently at Navier-Stokes level.
The theory is then applied to the modified-equilibrium model of Swift et al.\ \cite{swift_lattice_1995, swift_lattice_1996}. We find that, for this model, the noise must in general be spatially correlated to ensure thermalization at all length scales.
Additionally, we demonstrate that, at least for certain regions in the parameter space, satisfactory results at large length scales can also be achieved using an approximate, spatially uncorrelated form of noise derived in the hydrodynamic limit.
Finally, it is shown that capillary fluctuations can successfully be simulated using uncorrelated noise in the modified-equilibrium model.

\section{Continuum theory}
We first review the general physical background of the non-ideal fluid models considered here.
In this work, the following convention for the spatial and temporal Fourier transform of a quantity $a(\rv,t)$ is applied: $a(\kv,\omega) = (1/2\pi)^{d/2} \int d\rv dt\, a(\rv,t) \exp\left(\im(\kv\cdot \rv - \omega t)\right)$, where $d$ is the spatial dimension.
\subsection{Thermodynamics}
\begin{figure}[t]
\centering
    (a)\includegraphics[width=0.55\linewidth]{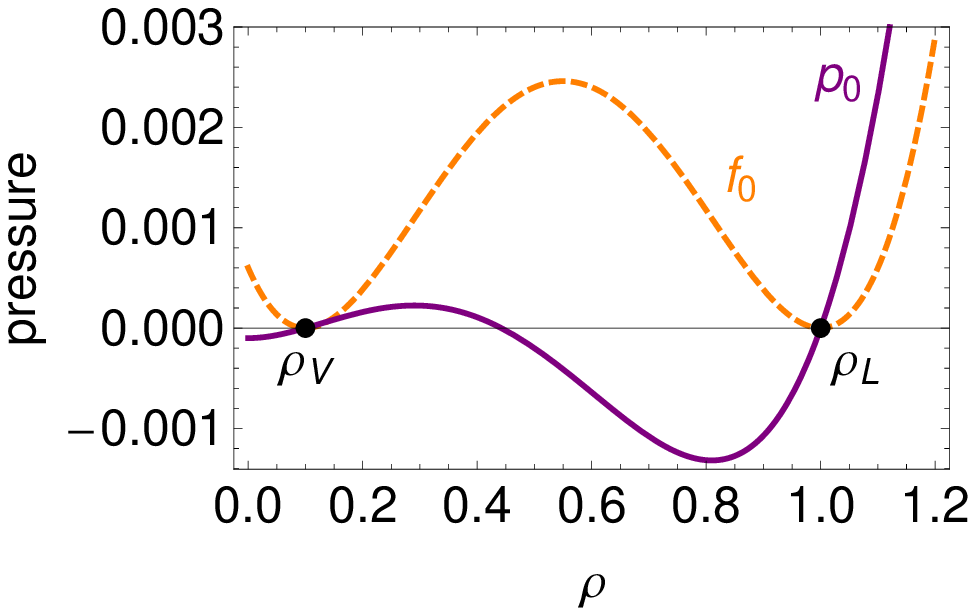}\\
    (b)\includegraphics[width=0.55\linewidth]{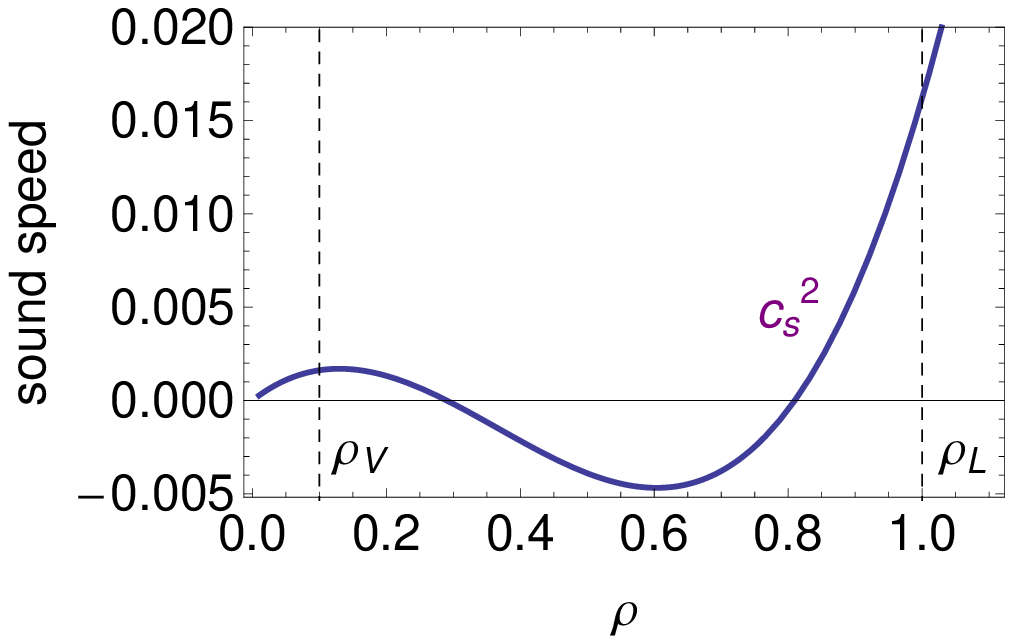}
   \caption{(Color online) Thermodynamic model: (a) bulk pressure $p_0$ and shape of the free energy density $f_0$ (not to scale), (b) sound speed squared. Parameters: $\rho_V = 0.1$, $\rho_L=1.0$, $\beta=0.01$.}
    \label{fig:bulk-prop}
\end{figure}

Our treatment is based on non-ideal fluid models described by a square-gradient free energy functional,
\beq
\mathcal{F}[\rho] = \int dV\left(f_0(\rho) + \frac{\kappa}{2}(\nabla \rho)^2 \right)\,.
\label{eq:freeE-funct}
\eeq
Here, $f_0$ is the bulk free energy density and $\kappa$ is the `square-gradient' parameter, which can be related to the surface tension and interface width. For $f_0$, we take a simple Landau-type double-well potential \cite{RowlinsonWidom_book, Chaikin_book, Jamet_Second_2001}:
\beq f_0(\rho) = \beta (\rho - \rho_V)^2(\rho-\rho_L)^2
\,,
\label{eq:f0-bulk}
\eeq
where $\rho_{V,L}$ are the desired equilibrium vapor and liquid densities and the parameter $\beta$ is inversely proportional to the compressibility of the fluid. The bulk pressure $p_0$ (equation of state) can be computed from the free energy density via $p_0 = \rho \partial_\rho f_0 - f_0$. Fig.~\ref{fig:bulk-prop} illustrates the typical shape of the bulk free energy, bulk pressure and speed of sound, $c_s^2 = \partial p_0 / \partial \rho = \rho \partial^2 f_0 / \partial\rho^2$.

In a single phase, the parameter $\beta$ is related to the speed of sound $c_s$ by
$$\beta = \frac{c_s^2}{2\rho_0 (\rho_L-\rho_V)^2}\,,$$
where $\rho_0$ corresponds to either $\rho_L$ or $\rho_V$, depending on which phase $c_s$ is referring to.
Typical values of $c_s$ in our simulations range from $0.04$ to $0.3$ in lattice units (l.u.) (see section \ref{sec:results}), hence the compressibility of the simulated non-ideal fluid is strongly enhanced compared to the ideal gas case, where $c_{s,\text{ideal}}=\sqrt{1/3}\simeq 0.57$ (Fig.~\ref{fig:bulk-prop}).
Coexisting phases in a square-gradient fluid are generally separated by a diffuse interface \cite{anderson_diffuse_1998, RowlinsonWidom_book}, which---for the above form of the free energy potential---has a width of
\beq \xi = \sqrt{\frac{8\kappa}{\beta}} \frac{1}{\rho_L - \rho_V} = 4\frac{\sqrt{\rho_0\,\kappa}}{c_s}\,.
\label{eq:interf-width}\eeq
The surface tension follows as
\beq \sigma = \frac{(\rho_L - \rho_V)^3}{6}\sqrt{2\kappa\beta} \,.
\label{eq:surf-tension}\eeq

The thermodynamic pressure tensor that follows from the above free energy functional is given by \cite{yang_molecular_1976, evans_interface_1979, goldstein_book, anderson_diffuse_1998, zou_multiphase_1999}
\beq P\ut{th} = \left(p_0 - \kappa \rho \nabla^2 \rho - \frac{\kappa}{2}|\nabla\rho|^2\right) \Imat + \kappa (\nabla \rho) \otimes (\nabla \rho)\,.
\label{eq:press-ten}
\eeq
Due to the square-gradient term, the pressure tensor receives non-local contributions in addition to the bulk pressure $p_0$.
In the Navier-Stokes equations, the divergence of this tensor appears, which can be written as the sum of a gradient of the thermodynamic bulk pressure and a force-like term:
\beq \nabla \cdot P\ut{th} = \nabla p_0 - \kappa \rho \nabla \nabla^2\rho\,.
\label{eq:thermo-force}
\eeq

Another route to derive the thermodynamic interaction force consists of directly computing the effective chemical potential from the free energy functional,
\beqn \mu = \frac{\delta \mathcal{F}}{\delta \rho} = \mu_0 - \kappa \nabla^2 \rho\,, \eeqn
with $\mu_0 = \partial_\rho f_0$.
The effective ``chemical'' body force that is acting on a fluid element is thus given by
$$\bv{F}\st{eff} = -\rho \nabla \mu = -\nabla p_0 + \kappa \rho \nabla \nabla^2 \rho = -\nabla\cdot P\st{th}\,.$$

\subsection{Fluctuations}
We consider fluctuations around a quiescent, homogeneous equilibrium state of density $\rho_0$ and vanishing macroscopic flow velocity $\uv_0=0$, i.e.\ $\rho(\rv)=\rho_0 + \delta\rho(\rv)$ and $\uv(\rv)=\delta \uv(\rv)$. The fluid momentum is given by $\delta \bv{j} = \rho_0 \delta \uv$.

\emph{Density fluctuations.} The essential difference between an ideal gas and a non-ideal fluid is the fact that, in the latter, density fluctuations are spatially correlated. The density correlation function (static structure factor) can be determined by expanding the free energy functional up to second-order in the density around equilibrium \cite{Hansen_ThoSL, Landau_StatPhys1, Chaikin_book}. By this, one obtains a Gaussian probability density in Fourier-space, with a variance given by an Ornstein-Zernike type structure factor
\beq  \bra \delta \rho(\kv) \delta \rho(\kv') \ket \equiv S(\kv)\delta(\kv+\kv') = \frac{\rho_0 k_B T}{c_s^2 + \rho_0 \kappa k^2} \delta(\kv+\kv') \,.
\label{eq:struct-fact}
\eeq
Here, $k_B$ the Boltzmann constant and $T$ is the temperature of the fluid.
The correlation length associated with the density fluctuations is given by $\sqrt{\rho_0 \kappa}/{c_s}$, which is directly proportional to the interface width, eq.~\eqref{eq:interf-width}.

\emph{Momentum fluctuations.} In an equilibrium fluid, the momenta of the fluid particles are always uncorrelated \cite{Hansen_ThoSL}, hence, by equipartition, the equal-time momentum correlation function is given by \cite{Landau_FluidMech59}
\beq \bra \delta j_\alpha(\kv) \delta j_\beta(\kv') \ket = \rho_0 k_B T \, \delta_{\alpha\beta} \,\delta(\kv+\kv')\,.\label{eq:momtm-fluct}\eeq

\subsection{Hydrodynamics}
\label{sec:hydro}
Using expression \eqref{eq:thermo-force} for the divergence of the pressure tensor for a fluid with an underlying square-gradient free energy functional, one obtains the linearized stochastic Navier-Stokes equations in $d$ dimensions as
\beq
\begin{split}
\partial_t \delta \rho &=- \rho_0 \nabla\cdot \bv{u}\,,\\
\rho_0 \partial_t \bv{u} &= -c_s^2 \nabla \delta \rho  + \kappa \rho_0 \nabla(\nabla^2 \delta\rho) \\
&+ \eta \nabla^2 \bv{u} + \left(\zeta+\eta[1-2/d]\right) \nabla(\nabla\cdot \bv{u}) - \nabla \cdot R\,.
\end{split}
\label{eq:fluct-nse}
\eeq
Here, $\eta$ and $\zeta$ are the shear and bulk viscosities and $R$ is the random stress tensor \cite{Landau_FluidMech59}.
We now Fourier transform the space dependence of $\delta \rho$, $\uv$ and $R$, and separate the velocity and the random stress tensor into longitudinal and transverse components, $\bv{u} = u_l \hat \kv + \bv{u}_t$, where $\hat\kv \equiv \kv/|\kv|$, $u_l \equiv \bv{u}\cdot \hat \kv$, and $\uv_t \equiv \bv{u}\cdot (\Imat - \hat \kv \hat \kv)$, and analogously   $\hat \kv\cdot R = R_l\hat \kv + \bv{R}_t$, where $R_l \equiv \hat \kv\cdot R \cdot \hat \kv$, and $\bv{R}_t \equiv \hat \kv\cdot R \cdot (\Imat-\hat\kv \hat\kv)$. We thus arrive at \cite{BoonYip_book, Reichl_book, Hansen_ThoSL}
\begin{align}
\partial_t \delta \rho &= \im\rho_0 k u_l \,, \label{eq:lin-ns1}\\
\partial_t u_l &= \im k \left( c_s^2 + \rho_0 \kappa k^2\right) \frac{\delta \rho}{\rho_0} - \nu_l k^2 u_l  + \frac{\im k}{\rho_0} R_l\,,\label{eq:lin-ns2}\\
\partial_t \uv_t &= -\nu_t k^2 \uv_t + \frac{\im k}{\rho_0}\bv{R}_t
\,,\label{eq:lin-ns3}
\end{align}
where we have introduced the longitudinal and transverse kinematic viscosities $\nu_l = [\zeta + \eta \left(2-2/d\right)]/\rho_0$, and $\nu_t = \eta/\rho_0$.

The essential observation is that the above equations are identical to those for a simple bulk fluid (i.e., a fluid where $\kappa=0$), if one introduces a wavenumber dependent sound speed
\beq
c_s^2(\kv) \equiv c_s^2 + \rho_0 \kappa k^2 = \rho_0 k_B T / S(\kv)\,
\label{eq:sound-speed-k}
\eeq
in the latter.
Further, we note that also for a square-gradient fluid it remains true that density fluctuations only couple to longitudinal momentum fluctuations, while transverse momentum fluctuations are completely decoupled from the other variables.

The fluctuation-dissipation theorem relates the equal-time correlation function of the fluid momentum to the correlation function of the fluctuating stress tensor $R$, which is assumed to obey a Gaussian probability distribution and have a Markovian character, $\bra R_{\alpha\beta}(\rv_1,t_1)R_{\alpha\beta}(\rv_2,t_2)\ket = A_{\alpha\beta}(\rv_1-\rv_2)\delta(t_1-t_2)$, with a variance $A_{\alpha\beta}$ to be specified below.
Since the non-ideal fluid interactions enter the hydrodynamic equations only through a modified speed of sound, one might expect that the expressions of the fluctuating stress tensor of a square-gradient fluid and a simple bulk fluid are identical. This is in fact true \cite{KimMazenko_1991}, and it holds even in the case of non-linear fluctuating hydrodynamics, if one properly takes into account the local values of the transport coefficients in the random stress tensor (which then represents multiplicative noise).

In the following, we explicitly demonstrate this fact and derive the expression for the variance of $R$ within the Langevin framework.
Inserting the continuity equation \eqref{eq:lin-ns1} into eq.~\eqref{eq:lin-ns2} and solving for the longitudinal velocity, one obtains
\beqn \partial_t^2 u_l = -k^2 c_s^2(k) u_l - \nu_l k^2 \partial_t u_l + \frac{\im k}{\rho_0} \partial_t R_l
\,.
\eeqn
Fourier transforming in time, we obtain the linear response relation:
\begin{multline}
u_l(\kv,\omega) = \frac{\omega k}{\rho_0 \left(\omega^2 - k^2 c_s^2(k) - \im\omega \nu_l k^2\right)} R_l(\kv,\omega) \\\equiv \chi_l(\kv,\omega) R_l(\kv,\omega) \,,
\end{multline}
keeping in mind that the longitudinal susceptibility $\chi_l(\omega)$ has poles at complex frequencies.
Squaring this equation, averaging over the noise and employing the white-noise property of $R$, $\bra |R_{\alpha\beta}(\kv,\omega)|^2\ket = A_{\alpha\beta}(\kv) $, the equal-time correlation function of $u_l$ follows after an inverse Fourier transform as
\begin{multline} \bra |u_l(\kv,t=0)|^2 \ket = \frac{1}{2\pi} \int d\omega \bra |u_l(\kv,\omega)|^2 \ket \\= \frac{A_l(\kv)}{2\pi}\int d\omega |\chi_l(\kv,\omega)|^2
\,.
\label{eq:vel-noise-avg}
\end{multline}
Here, we introduced the longitudinal component $A_l$ of the variance $A_{\alpha\beta}$.
The integral over $\chi_l(\omega)$ can be computed using contour integration, giving $\int d\omega |\chi_l(\omega)|^2 = \pi / \rho_0^2 \nu_l$.
In order to obtain a thermally equilibrated fluid, the noise average of the velocity correlator in \eqref{eq:vel-noise-avg} is set equal to the thermal average.
Since equipartition demands
$\bra |u_l(\kv)|^2 \ket = k_B T /\rho_0\,,$
we finally obtain \footnote{We neglect here any dimensional factors due to delta functions $\delta(0)$ evaluated at zero.}:
\beq \bra |R_l(\kv,\omega)|^2\ket = 2 k_B T \rho_0 \nu_l\,.
\label{eq:ns-random-stress-l}
\eeq
The corresponding relation for each transverse component of $R$ follows analogously as
\beq \bra |R_t(\kv,\omega)|^2\ket = 2 k_B T \rho_0 \nu_t\,.
\label{eq:ns-random-stress-t}
\eeq
Hence, the ``classical'' FDT of fluctuating hydrodynamics for a simple bulk fluid is recovered, with a random stress tensor that is uncorrelated in space and time. Fourier-transforming back to the real space and time domain, the FDT assumes the well-known form \cite{Landau_FluidMech59, Reichl_book}
\begin{widetext}\beq \bra R_{\alpha \beta}(\bv{r},t) R_{\gamma \delta}(\bv{r'},t') \ket = 2k_B T \left[ \eta \left(\delta_{\alpha \gamma} \delta_{\beta \delta} + \delta_{\alpha \delta} \delta_{\beta \gamma} - \frac{2}{d} \delta_{\alpha \beta}\delta_{\gamma \delta}\right)
+ \zeta\, \delta_{\alpha \beta}\delta_{\gamma \delta}\right]\delta(\bv{r}-\bv{r'})\delta(t-t')
\,.
\label{eq:fluct-hydro-fdt}\eeq
\end{widetext}
Note that the last expression is valid only if the viscosities are independent of $k$, as is the case for a simple bulk fluid. The particular non-ideal fluid LB model we consider in section \ref{sec:yeo-model}, however, entails a $k$-dependent bulk viscosity [see eq.~\eqref{eq:bulkv-yeo}] and hence the random stress tensor becomes spatially correlated in this case.

In an \emph{inhomogeneous} state, i.e., in the presence of interfaces, the fluid is governed by non-linear equations of motion since the non-linear terms originating from the pressure tensor \eqref{eq:press-ten} can not be neglected anymore. However, small fluctuations around a solution of the full non-linear equations can still be locally described by the same linearized equations of motion \eqref{eq:fluct-nse} as in the uniform case.
Therefore, the FDT of fluctuating hydrodynamics \eqref{eq:fluct-hydro-fdt} is expected to remain valid if the effect of the inhomogeneity is taken into account in the local values of the thermodynamic quantities and the transport coefficients \cite{tremblay_noneq_1981, KimMazenko_1991}.
This renders the noise effectively multiplicative \cite{risken_fp_book}.
Applying the linear Langevin formalism to general non-equilibrium situations can often be justified along similar arguments after a local-equilibrium assumption has been made \cite{tremblay_noneq_1981, keizer_fluct_1978}.

%----------------------------------------------------------------------------------------

\section{Lattice Boltzmann modeling}
We now turn to the LB model that corresponds to the continuum theory of non-ideal fluids presented in the preceding section.
Although we focus on a two-dimensional system, all derivations are kept as general as possible and are easily applied to three dimensions.
The fluctuating LBE is most conveniently developed in terms of moments of the distribution function. The moment space is divided into the conserved, transport and ghost (or kinetic) sectors. The FDT is derived below in Fourier-space by treating all LB modes on an equal footing, which, as first pointed out in \cite{adhikari_fluct_2005}, is a necessary prerequisite to achieve complete thermalization of the fluid.
The evaluation of the FDT requires, besides information on the relaxation and interaction behavior (which is provided by the LB model itself), additional information in the form of the correlation matrix of the LB modes. The latter ingredient has to be derived from a statistical mechanical framework. The FDT is finally applied to the modified-equilibrium non-ideal fluid model of Swift et al.~\cite{swift_lattice_1995, swift_lattice_1996}.

\emph{Notation and conventions.} The spatial Fourier transform of a quantity $a(\rv)$ defined on the lattice is computed according to
$a(\bv{k})= \frac{1}{\sqrt{n}} \sum_\bv{r} e^{\im \bv{k}\cdot \bv{r}} a(\bv{r})\,,$
where $n$ is the total number of lattice points in the system. Since $a$ is usually a real quantity, it is sufficient to consider just the first quadrant of the first Brillouin zone, $k_\alpha=0\ldots \pi$, where $k_\alpha=\pi$ corresponds to a physical length of 2 l.u. Fourier transforming derivative operators on a lattice requires additional care, as described in appendix \ref{sec:fourier-latt}.
In general, Greek indices refer to Cartesian coordinates, while Latin indices refer to the lattice directions or the LB moments. Repeated free indices are to be summed over.
In the following, the quantity $\cslb\equiv \sqrt{1/3}$ is a constant specific to the chosen lattice and agrees with the speed of sound of the ideal LB gas. It has to be distinguished from the actual speed of sound $c_s$ of a non-ideal fluid.

\subsection{Introduction}
We begin by reviewing the necessary theory of the deterministic LBE, which is given by
\beq
 f_i(\rv +\bv{c}_i, t+1) = f_i+ \Lambda_{ij}(f_j - f_j\ueq) + F_i\,.
\label{eq:gen-lbe}
\eeq
Note that the $\rv$- and $t$-dependences have been suppressed on the right hand side of the equation.
Here, $F_i$ describes a possible body force and $\Lambda_{ij}$ is a general matrix relaxation operator \cite{higuera_epl1989, higuera_succi_epl1989}. The equilibrium distribution $f_i\ueq$ is model-dependent and will be specified later. In this work, we consider a D2Q9 lattice, hence $i=1,\ldots,9$.
For a general discussion of the LB method for simple and complex fluids, we refer to the literature \cite{succi_book, nourgaliev_lbm_2003, raabe_overview_2004, benzi_physrep1992, aidun_rev2010}.

For the present purposes, it proves to be most convenient to work in the space of moments $m_a$ ($a=1,\ldots,9$) of the distribution function $f_i$ \cite{dHumieres_MRT_1992}. This can be achieved by constructing a set of orthogonal basis vectors $T_{ai}$ from the lattice velocities $\cv_i$. Orthogonality is measured with respect to the weighted scalar product \cite{adhikari_fluct_2005},
\beq \bra T_a | T_b \ket \equiv  w_i T_{ai} T_{bi} = N_a \delta_{ab},
\label{eq:scalar-prod}
\eeq
where $N_a$ is the length of the $a$th basis vector $T_a$, $N_a= \sum_i w_i T_{ai}^2$. The weights $w_i$ are identical to the ones used in the definition of the (ideal gas) equilibrium distribution. For a standard D2Q9 lattice, these are
$$w_1 = 4/9,\quad w_{2\ldots 5} = 1/9,\quad w_{6\ldots 9} = 1/36 \,.$$
With this choice, the projection of the (ideal gas) equilibrium distribution onto the ghost modes is eliminated \cite{adhikari_fluct_2005}. Although this is in principle not necessary for the present developments, we will adopt this choice henceforth, since it emphasizes the physical background of the model and is computationally advantageous.

The moments are now defined as the projections of the distribution function onto the basis vectors,
\beqn m_a =  T_{ai} f_i\,.
\eeqn
In turn, any vector defined in velocity space (such as the distribution function $f_i$) can be expanded in terms of the orthogonal basis vectors,
\beqn f_i = (T^{-1})_{ia}m_a = w_i  T_{ai} m_a/N_a\,.
\eeqn
In this relation, the expression for the inverse transformation matrix, $(T^{-1})_{ia} = w_i T_{ai} /N_a$, has been used.

Once a suitable basis set is chosen (see below), one constructs a collision operator $\Lambda$ that is diagonal in moment space by setting $\Lambda = T^{-1} \hat \Lambda T$, where $\hat \Lambda = \mathrm{diag}(\lambda_{a=1,\ldots,9})$ is a diagonal matrix of relaxation parameters $\lambda_a$.
Hence, the basis vectors $T_{a}$ are eigenvectors of the generalized collision operator $\Lambda$ with eigenvalues $\lambda_a$.
The $\lambda_a$ can be expressed in terms of relaxation times $\tau_a$ through $\lambda_a = -1/\tau_a\,.$ Well-known stability requirements impose the restriction $\tau_a>1/2$ \cite{succi_book}.
Rewriting the right hand side of eq.~\eqref{eq:gen-lbe} in terms of moments, the LBE becomes
\beq
f_i(\bv{x}+\bv{c}_i , t+1) = T_{ia}^{-1}\left[ m_a + \lambda_a (m_a - m_a\ueq) + m^F_a  \right]\,,
\label{eq:gen-lbe-m}
\eeq
where $m^F_a = T_{ai} F_i$ are the moments of the forcing term.

The D2Q9 basis set used in the present work is summarized in Table \ref{tab:modes-d2q9} \cite{duenweg_statmechLB_2007}.
The first three rows cover the conserved hydrodynamic moments, i.e.\ the density and momentum
\beqn \rho = \sum_i f_i\,,\quad \bv{j} = \sum_i \cv_i f_i\,.
\eeqn
The next three rows contain the non-conserved hydrodynamic (transport) moments, and the last three contain the ghost (or kinetic) moments.
\begin{table}[t]
\begin{center}
\begin{tabular}{r | c | c | c | c}
$a$ & $T_{ai}$ & $N_a$ & $m_{a}$ & $\lambda_a$ \\
\hline
 1 & 1                                    &  1   & $\rho$   & 0 \\
 2 & $c_{ix}$                             & 1/3  & $j_x$    & 0\\
 3 & $c_{iy}$                             & 1/3  & $j_y$    & 0\\
 \hline
 4 & $3c_i^2-2$                           &  4   & $e$      & $\lambda_b$ \\
 5 & $2c_{ix}^2-{c}_i^2$                  & 4/9  & $p_{ww}$    & $\lambda_s$ \\
 6 & $c_{ix} {c}_{iy}$                    & 1/9  & $p_{xy}$     & $\lambda_s$ \\
 \hline
 7 & $(3c_i^2-4){c}_{ix}$                 & 2/3  & $q_x$    & $\lambda_q$ \\
 8 & $(3c_i^2-4){c}_{iy}$                 & 2/3  & $q_y$    & $\lambda_q$ \\
 9 & $9c_i^4-15{c}_i^2+2$                 & 16   & $\epsilon$   & $\lambda_\epsilon$ \\
\hline
\end{tabular}
\end{center}
\caption{Basis set of the D2Q9 model. $T_{ai}$ denotes the basis vector, $N_a$ its length, $m_{a}$ is the designation of the corresponding moment and $\lambda_a$ is its eigenvalue in the collision operator.}
\label{tab:modes-d2q9}
\end{table}
The moment $e$ describes a pressure or bulk stress mode, with an eigenvalue $\lambda_b$ related to the bulk viscosity.
$p_{ww}$ and $p_{xy}$ are shear modes, with a common eigenvalue $\lambda_s$ related to the shear viscosity.
The ghost modes consist of a ghost density mode $\epsilon$ and ghost vector current $q_\alpha$, in line with the duality prescription \cite{adhikari_duality_2008}.
Using the numerical expressions for the lattice velocities, the transformation matrix for the present basis set reads
\beq T = \left(
\begin{array}{ccccccccc}
 1 & 1 & 1 & 1 & 1 & 1 & 1 & 1 & 1 \\
 0 & 1 & 0 & -1 & 0 & 1 & -1 & -1 & 1 \\
 0 & 0 & 1 & 0 & -1 & 1 & 1 & -1 & -1 \\
 -2 & 1 & 1 & 1 & 1 & 4 & 4 & 4 & 4 \\
 0 & 1 & -1 & 1 & -1 & 0 & 0 & 0 & 0 \\
 0 & 0 & 0 & 0 & 0 & 1 & -1 & 1 & -1 \\
 0 & -1 & 0 & 1 & 0 & 2 & -2 & -2 & 2 \\
 0 & 0 & -1 & 0 & 1 & 2 & 2 & -2 & -2 \\
 2 & -4 & -4 & -4 & -4 & 8 & 8 & 8 & 8
\end{array}
\right)
\,.
\label{eq:Tmat}
\eeq

\subsection{FDT}
The fluctuating LBE is obtained from the deterministic LBE, eq.~\eqref{eq:gen-lbe-m}, by adding random noise variables $\xi_a$ to the collision step.
A small fluctuation around a uniform, global equilibrium state of density $\rho_0$ and vanishing flow velocity, $\delta f_i(\rv,t) = f_i(\rv,t) - f_i\ueq(\rho_0, u=0)$, or equivalently, $\delta m_a(\rv,t) = m_a(\rv,t) - m_a\ueq(\rho_0, u=0)$,
then evolves according to the linearized equation
\begin{multline}
 \delta f_i(\rv +\bv{c}_i, t+1) = T_{ia}^{-1}[ (1 + \lambda_a) \delta m_a - \lambda_a\delta m_a\ueq \\+ \delta m^F_a + \xi_a ]\,.
\label{eq:fluct-lin-lbe}
\end{multline}
The noise is assumed to be uncorrelated in time and drawn from a Gaussian probability distribution, whose covariance matrix can be expressed as $\bra \xi_a(\rv,t) \xi_b(\rv',t') \ket = \Xi_{ab}(\rv-\rv') \delta_{t,t'}$, assuming translational invariance. It further is assumed that the $\xi_a$ are uncorrelated with the LB modes $\delta m_a$. The matrix $\Xi_{ab}$ will be determined below by means of the fluctuation-dissipation theorem.

In order to rewrite the fluctuating LBE \eqref{eq:fluct-lin-lbe} fully in terms of moments, we apply a spatial Fourier transform and introduce the Fourier-transformed advection operator in moment space by \cite{lallemand_theory_2000}
\beq
A_{ab}(\kv) = T_{aj} \exp(-\im \bv{k}\cdot \bv{c}_j) (T^{-1})_{jb}\,.
\label{eq:lb-adv}
\eeq
This operator has the important property that $A^{-1}_{ab}(\kv)= A_{ab}^\star(\kv)\,$, where $^\star$ denotes complex conjugation.
Collecting the effect of relaxation and interactions in a linearized collision operator $\Omega(\kv)$, we finally obtain
\beq
\delta m_a(\bv{k},t+1) = A_{ab}(-\kv) \left[(\Imat+\Omega)_{bc}\delta m_c(\kv,t) + \xi_b(\bv{k},t) \right]\,.
\label{eq:lb-lang-ft}
\eeq
The computation of $\Omega$ requires only the knowledge of $\delta m\ueq$ and $\delta m^F$ and will be performed in section \ref{sec:yeo-model} for a specific LB model.

The derivation of the FDT for a non-ideal fluid model proceeds analogously to \cite{dufty_lblangevin_1993, adhikari_fluct_2005}.
We multiply eq.~\eqref{eq:lb-lang-ft} with $\delta m_d(-\kv,t+1)$ from the right, average over the noise distribution and assume stationarity of equal time correlators.
Introducing the equal-time correlation matrix of the modes as $G_{ab}(\kv)\equiv \bra \delta m_a(\kv) \delta m_b(-\kv)\ket$ and the (Fourier-transformed) covariance matrix of the noise as $\Xi_{ab}(\kv)\equiv \bra \xi_a(\kv) \xi_b(-\kv)\ket$, we obtain the FDT in the form
\beq
\Xi(\kv) = A(\kv) G(\kv) A(-\kv)^{T} - \big[\Imat +\Omega(\kv)\big]G(\kv)\big[\Imat + \Omega(-\kv)\big]^T
\label{eq:noise-matrix}
\,.
\eeq
Note that, due to the absence of non-linearities, the above expression can be independently evaluated for each point in $k$-space.

The equilibrium correlations of the modes, represented by the matrix $G$, are not immediately provided by the LB scheme, but must instead, as for any Langevin equation, be derived with the help of a statistical mechanical theory \cite{adhikari_fluct_2005, duenweg_statmechLB_2007, dufty_lblangevin_1993}.
A suitable ansatz for a general non-ideal fluid is provided by the following relation for the equal-time correlations of fluctuations in the distribution function:
\beq
\bra \delta f_i(\kv) \delta f_j(\kv') \ket = \Big[\bar f_i \bar f_j [S(\kv)/\rho_0-\mu]/\rho_0 + \mu \bar f_i \delta_{ij}\Big] \delta_{\kv,-\kv'}\,.
\label{eq:f-correl-lb}
\eeq
Here, $\bar f_i = f\ueq_i(\rho_0, u=0)$ denotes the global Maxwellian of the uniform reference state and $S(\kv)$ is the structure factor of the non-ideal fluid.
The parameter $\mu$ can be interpreted as the mass of a fictitious fluid particle and must be determined such that eq.~\eqref{eq:f-correl-lb} leads to the correct expression \eqref{eq:momtm-fluct} of the equal-time momentum correlations.
The correlation matrix $G$ is obtained from \eqref{eq:f-correl-lb} through a basis transformation,
\beq \begin{split}
G_{ab}(\kv) &= T_{ai} T_{bj} \bra \delta f_i(\kv) \delta f_j(-\kv) \ket \\
&= \bar m_a \bar m_b [S(\kv)/\rho_0-\mu]/\rho_0 + \mu T_{ai} T_{bi} \bar f_i \,,
\label{eq:g-correl}
\end{split}\eeq
where $\bar m_a = T_{ai} \bar f_i$.
A derivation of relation \eqref{eq:f-correl-lb} from continuum kinetic theory is presented in appendix \ref{sec:f-correl}.
The above relation is expected to be appropriate to non-ideal fluid models that are based on the ideal gas equilibrium distribution, which is the LB equivalent of the Maxwellian distribution used in continuum kinetic theory.
Conversely, non-ideal fluid models employing a non-standard equilibrium can be expected to require a different ansatz from \eqref{eq:f-correl-lb}.
This is indeed the case for the modified-equilibrium model considered below [see eq.~\eqref{eq:f-correl-yeo}].
The general theoretical status of \eqref{eq:f-correl-lb} in the context of the LB method for non-ideal fluids will be further investigated in future works.

In the hydrodynamic regime, i.e.\ at small $k$, the variances of the noise variables pertaining to the stress modes (here $\xi_{4,5,6}$) can be determined directly using the FDT of fluctuating hydrodynamics (appendix \ref{sec:fdt-hydro-lim}). This fact allows for an independent check of the noise constructed on LB level, eq.~\eqref{eq:noise-matrix}.
In many cases it turns out that satisfactory simulation results can already be achieved by using noise evaluated for $k \ra 0$, which is then spatially \emph{uncorrelated} by construction, similarly to the ideal gas case.
This behavior can be expected to apply whenever the non-ideal fluid interactions enter the dynamic equations in a fully reversible way---and hence, not give additional contributions to the dissipative terms---or if the irreversible contribution is sufficiently weak.
Since $A=\Imat$ at $k = 0$, the noise covariance matrix reduces in the zero wavelength-limit to
\beq \Xi(\bv{0}) \equiv \lim_{k \ra 0} \Xi(\kv) = - G \Omega^T - \Omega G - \Omega G \Omega^T
\,, \label{eq:noise-matrix-k0}\eeq
where all quantities on the right hand side are evaluated for $k \ra 0$.
This uncorrelated form of the noise can be constructed independently on each lattice site in real space.

Moreover, as remarked in section \ref{sec:hydro}, one can usually assume the FDT of fluctuating hydrodynamics \eqref{eq:fluct-hydro-fdt}, derived from the linearized Langevin equations, to remain still valid even for an \emph{inhomogeneous} fluid, if the local values of the thermodynamic quantities and the transport coefficients are taken into account in the computation of the noise. This implies, that uncorrelated noise described by $\Xi(\bv{0})$ can be readily applied to inhomogeneous systems, as will be further explained in the next section.

\subsection{Application to a non-ideal fluid model}
\label{sec:yeo-model}
\begin{table}[b]
\begin{center}
\begin{tabular}{r | c | c }
$a$ & \text{mode} & $m_a\ueq$ \\
\hline
 1 & $\rho$   & $\rho$ \\
 2 & $j_x$    & $\rho u_x$\\
 3 & $j_y$    & $\rho u_y$\\
 \hline
 4 & $e$      & $3\rho(u_x^2 + u_y^2)+ 6(p_0 - \rho\cslb^2 - \kappa \rho \nabla^2\rho) + C_1$\\
 5 & $p_{ww}$    & $\rho(u_x^2 - u_y^2) + \kappa \left[(\partial_x \rho)^2 - (\partial_y \rho)^2\right]+C_2$ \\
 6 & $p_{xy}$     & $\rho u_x u_y + \kappa (\partial_x \rho) (\partial_y \rho) +C_3$\\
 \hline
 7 & $q_x$    & 0 \\
 8 & $q_y$    & 0 \\
 9 & $\epsilon$   & $- 6(p_0 - \rho\cslb^2 - \kappa \rho \nabla^2\rho) -3\kappa \left[(\partial_x\rho)^2 + (\partial_y\rho)^2\right]$ \\
\hline
\end{tabular}
\end{center}
\caption{Equilibrium moments of the modified-equilibrium non-ideal fluid LB model. The $C_n$ denote Galilean-invariance correction terms.}
\label{tab:mom-yeo}
\end{table}
We now apply the general FDT derived above to the modified equilibrium model proposed by Swift et al.~\cite{swift_lattice_1995, swift_lattice_1996, pooley_spurious_2008}. In this approach, the non-ideal fluid interactions are derived from a square-gradient free energy functional, eq.~\eqref{eq:freeE-funct}, and hence, the stochastic version of this model is supposed to approximate the fluctuating Navier-Stokes equations \eqref{eq:fluct-nse} on large length and time scales.
The non-ideal fluid interactions enter this model through a stress contribution to a modified equilibrium distribution whose moments are stated in Table~\ref{tab:mom-yeo}. The $C_n$ are correction terms that ensure Galilean-invariance of the model up to O(Ma$^2$) \cite{holdych_gal_1998}. Since they are at least of second order in $\rho$ und $\uv$, their specific form is not important for the linearized model. We notice, that, in contrast to the ideal gas model, the equilibrium distribution has a non-vanishing projection onto a ghost mode. The evolution equation for the modified-equilibrium model is given by eq.~\eqref{eq:gen-lbe-m}, without a forcing term (i.e.\ $m^F_a=0$).

In the linearized model, after applying a spatial Fourier transform (see appendix \ref{sec:fourier-latt}) and writing $\delta p_0 = c_s^2 \delta\rho$, the fluctuations of the equilibrium modes follow from Table~\ref{tab:mom-yeo} as
\begin{multline}\delta m_a\ueq = \{\delta \rho, \delta j_x, \delta j_y,
 d(\kv)\delta \rho, 0, 0, 0, 0, -d(\kv)\delta \rho\}\,,
\label{eq:eq-mom-yeo} \end{multline}
where we invoked the definition of the generalized speed of sound, $c_s^2(\kv) = c_s^2 + \rho_0 \kappa k^2$, and defined $d(\kv)\equiv  6\left[c_s^2(\kv) -\cslb^2\right]$.
The model can eventually be brought into the general form of eq.~\eqref{eq:lb-lang-ft} by introducing the matrix collision operator as
\beq
\Omega(\kv) = \left(
\begin{array}{ccc|ccc|ccc}
 . & . & . & . & . & . & . & . & . \\
 . & . & . & . & . & . & . & . & . \\
 . & . & . & . & . & . & . & . & . \\
\hline
 -\lambda_b d(\kv) & . & . & \lambda_b & . & . & . & . & . \\
 . & . & . & . & \lambda_s & . & . & . & . \\
 . & . & . & . & . & \lambda_s & . & . & . \\
\hline
 . & . & . & . & . & . & \lambda_q & . & . \\
 . & . & . & . & . & . & . & \lambda_q & . \\
 \lambda_\epsilon d(\kv) & . & . & . & . & . & . & . & \lambda_{\epsilon} \\
\end{array}
\right),
\label{eq:coll-op-yeo}
\eeq
where the dots indicate zeros for short.

To compute the equilibrium correlation matrix $G=\bra \delta m_a \delta m_b^*\ket = T_{ai} T_{bj} \bra \delta f_i \delta f_j^* \ket$ for the modified-equilibrium model, we propose here the ansatz
\beq
\bra \delta f_i(\kv) \delta f_j(-\kv) \ket = \frac{S(\kv)}{\rho_0} \bar f_i(\kv) \delta_{ij}\,,
\label{eq:f-correl-yeo}
\eeq
where $S(\kv)$ is the structure-factor, defined by eq.~\eqref{eq:struct-fact}, taking into account the proper lattice derivative operators (see appendix \ref{sec:fourier-latt}), and $\bar f_i(\kv)\equiv f_i\ueq(\kv,u=0)$ is the equilibrium distribution of the model evaluated for vanishing flow velocity.
This expression can be interpreted as the natural generalization of the corresponding ideal gas relation, $\bra \delta f_i \delta f_j\ket = S\st{id}\bar f_i \delta_{ij}/\rho_0 $, to the present non-ideal fluid model. The connection of relation \eqref{eq:f-correl-yeo} to the continuum kinetic theory of non-ideal fluids and further motivations pointing to its validity are detailed in appendix \ref{sec:f-correl}.
In moment space, the correlation matrix becomes
%\vspace{2cm}
\begin{widetext}
\beq
G(\kv) = \left(
\begin{array}{ccc|ccc|ccc}
 S(\kv) & . & . & S(\kv)d(\kv) & . & . & . & . & - S(\kv)d(\kv) \\
 . & \tilde T \cslb^2 & . & . & . & . & . & . & . \\
 . & . & \tilde T \cslb^2 & . & . & . & . & . & . \\
\hline
 S(\kv)d(\kv) & . & . & N_4 S(\kv) & . & . & . & . & 2 S(\kv)d(\kv) \\
 . & . & . & . & N_5 \tilde T & . & . & . & . \\
 . & . & . & . & . & N_6 \tilde T & . & . & . \\
\hline
 . & . & . & . & . & . & N_7\tilde T & . & . \\
 . & . & . & . & . & . & . & N_8\tilde T & . \\
 -S(\kv)d(\kv) & . & . & 2 S(\kv)d(\kv) & . & . & . & . & N_9 (S(\kv)+3\tilde T)/4  \\
\end{array}
\right),
\label{eq:g-mat-yeo}
\eeq
%\end{widetext}
where the $N_a$ are the lengths of the basis vectors and \mbox{$\tilde T \equiv\rho_0 k_B T/\cslb^2$} for short.

Evaluating the noise covariance matrix \eqref{eq:noise-matrix} using the expressions for $\Omega(\kv)$, eq.~\eqref{eq:coll-op-yeo}, and $G(\kv)$, eq.~\eqref{eq:g-mat-yeo}, shows that the advective contribution cancels, i.e.\ $AGA^\dagger=G$ for all $\kv$, just as in the ideal gas model \cite{adhikari_fluct_2005} and in the continuum Boltzmann-equation \cite{klimontovich_nonideal_1973}. The noise covariance matrix is finally obtained as
%\begin{widetext}
\beq
\Xi(\kv) = -\frac{\rho_0 k_B T}{\cslb^2} \left(
\begin{array}{ccc|ccc|ccc}
 . & . & . & . & . & . & . & . & . \\
 . & . & . & . & . & . & . & . & . \\
 . & . & . & . & . & . & . & . & . \\
\hline
 . & . & . & N_4 \left[2-3c_s^2(\kv)\right] \tilde \lambda_e & . & . & . & . & 12 \left[c_s^2(\kv)-\cslb^2\right] \tilde \lambda_{e\epsilon} \\
 . & . & . & . & N_5 \tilde \lambda_s & . & . & . & . \\
 . & . & . & . & . & N_6 \tilde \lambda_s & . & . & . \\
\hline
 . & . & . & . & . & . & N_7\tilde \lambda_q & . & . \\
 . & . & . & . & . & . & . & N_8\tilde \lambda_q & . \\
 . & . & . & 12 \left[c_s^2(\kv)-\cslb^2\right] \tilde \lambda_{e\epsilon} & . & . & . & . & N_9\left[\frac{5}{4} - \frac{3}{4} c_s^2(\kv)\right]\tilde \lambda_\epsilon  \\
\end{array}
\right),
\label{eq:noise-yeo}
\eeq
\end{widetext}
where we defined $\tilde \lambda_a \equiv \lambda_a(2+\lambda_a)$ and $\tilde \lambda_{e\epsilon} \equiv \lambda_e + \lambda_\epsilon + \lambda_e\lambda_\epsilon$ for short.
This result shows that the modified-equilibrium model requires spatially correlated noise to satisfy the FDT at all scales.
This is in complete agreement with the FDT of fluctuating hydrodynamics (see appendix \ref{sec:fdt-hydro-lim}) for this model:
the bulk viscosity of the modified-equilibrium model is found to be given by
\beq \zeta(\kv) = \rho_0 \cslb^2 \left(\tau_b -\onehalf\right)\left[2-\frac{c_s^2(\kv)}{\cslb^2}\right]\,,\label{eq:bulkv-yeo}\eeq
and hence (in contrast to an ideal gas model) is wavenumber-dependent \cite{nourgaliev_lbm_2003}.
Indeed, we recognize in $\Xi_{44}$ the presence of the same correction factor $2-3c_s^2(\kv)$ that also appears in the bulk viscosity.
The shear viscosity is the same as in an ideal gas model, $\eta = \rho_0 \cslb^2 (\tau_s -\onehalf)$, in further agreement with result \eqref{eq:noise-yeo}.
We shall take these observations as crucial hints to the correctness of the ansatz \eqref{eq:f-correl-yeo} for the modified-equilibrium model.
Results \eqref{eq:noise-yeo} and \eqref{eq:bulkv-yeo} indicate that the non-ideal interactions are implemented in this model in a way that is not fully reversible.

A useful approximation to the full noise matrix consists of evaluating $\Xi$ in the limit $k \ra 0$, resulting in spatially uncorrelated noise. Although this form of noise satisfies the FDT of fluctuating hydrodynamics strictly only in the zero wavenumber-limit, simulations indicate that, at least for certain parameter ranges, satisfactory thermalization is obtainable also well in the finite $k$ regime.
This can be explained by the weak $k^2$-dependence of the generalized speed of sound $c_s(\kv)$, and hence of the noise \eqref{eq:noise-yeo} and the bulk viscosity \eqref{eq:bulkv-yeo}.
Noise defined by $\Xi(\b0)=\lim_{k \ra 0}\Xi(\kv)$ differs from the noise of an ideal gas model by the presence of cross-correlations between stress ($e$) and ghost ($\epsilon$) noises, which originate from the cross-correlations in the equilibrium correlation matrix \eqref{eq:g-mat-yeo}. For the simulation of hydrodynamics at small wavenumbers, these cross-correlations are immaterial as the ghost modes are decoupled from the hydrodynamic modes.

\begin{figure}[b]\centering
    \includegraphics[width=0.55\linewidth]{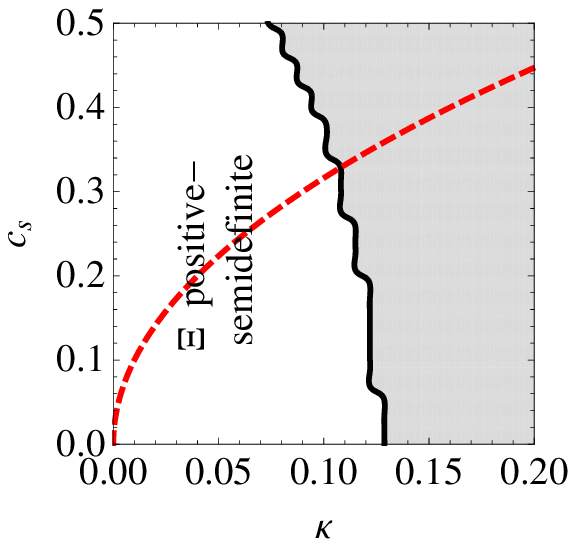}
   \caption{(Color online) Allowed region of simulation parameters. The white area corresponds to the combinations of $\kappa$ and $c_s$ that ensure a positive-semidefinite noise covariance matrix $\Xi(\kv)$ for all $\kv$. The dashed red curve marks the parameter combinations that result in an interface width of 4 lattice units. Points above the curve correspond to smaller interface widths and thus can lead to potentially unstable simulations when interfaces are present. (All relaxation times are set to 1.0.)}
    \label{fig:allowed-param}
\end{figure}

The requirement of positive-definiteness of $\Xi$ imposes a restriction on the allowed values of the square-gradient parameter $\kappa$ and the sound speed $c_s$ through their relation to the generalized speed of sound $c_s^2(\kv) = c_s^2 + \rho_0 \kappa k^2$ \footnote{$k^2$ has to be understood here as the negative of the Fourier-transformed Laplace operator, see appendix \ref{sec:fourier-latt}.}. This is demonstrated in Fig.~\ref{fig:allowed-param}, where the white region corresponds to parameter combinations that ensure a positive-semidefinite noise covariance matrix for all relevant wavenumbers ($k_\alpha=0,\ldots,\pi$). Note that this region in general also depends on the relaxation rates $\lambda_e$ and $\lambda_\epsilon$ (we have chosen $\lambda_e=\lambda_\epsilon=1$ in the plot), but it is independent of the temperature and density. The \emph{actually} allowed combinations of simulation parameters are a subset of the white region in Fig.~\ref{fig:allowed-param}, as numerical stability provides further restrictions. In particular, a lower bound of the interface width leads, via eq.~\eqref{eq:interf-width}, to an upper bound on the sound speed in the liquid phase for each value of $\kappa$ (dashed curve in Fig.~\ref{fig:allowed-param}).

We finally remark that the fluctuation temperature $T$, which is related to the velocity fluctuations by eq.~\eqref{eq:momtm-fluct}, is bounded from above by the stability constraint of LB (known as low Mach-number constraint in the ideal gas case), $k_B T/\rho_0 = \bra u_\alpha^2 \ket \ll \cslb^2 $, or
\beq k_B T \ll \cslb^2 \rho_0 \,. \label{eq:low-mach}\eeq
While this constraint also holds for simulations of thermal fluctuations in the ideal gas \cite{adhikari_fluct_2005}, it becomes particularly relevant in a two-phase system, as the above constraint must be fulfilled both in the liquid and gas phases with a uniform temperature throughout the system.
Thus, in a two-phase system, the maximal attainable temperature is limited by the vapor density.
In particular, one has $k_B T = \rho_L \bra u_{L,\alpha}^2\ket = \rho_G \bra u_{G,\alpha}^2\ket$, hence $\bra u_{L,\alpha}^2\ket = (\rho_G/\rho_L) \bra u_{G,\alpha}^2\ket$, showing that the effects of thermal fluctuations on the kinetics in the liquid phase reduce with increasing density ratio.
This implies furthermore, that systems with high density ratios are effectively simulated at correspondingly larger length scales, where fluctuations are less pronounced.

\section{Implementation}
We shall briefly describe here the practical steps required to implement thermal noise in a simulation based on the modified-equilibrium model. Although the theoretical analysis in the preceding section has been performed for an underlying D2Q9 lattice, the steps in the derivation of the FDT are nevertheless general and applicable to any D$d$Q$n$ lattice. The resulting noise covariance matrix $\Xi$ is expected to remain of similar form to \eqref{eq:noise-yeo}.

As mentioned in the preceding section, there exist in principle two options to model thermal fluctuations. The computationally easiest one is to employ spatially \emph{uncorrelated} noise, in which case the quantities $\xi_a(\rv)$ in the LBE \eqref{eq:fluct-lin-lbe} are spatially independent Gaussian random variables of covariance given by the zero wavenumber-limit $\Xi(\bv{0})= \lim_{k\ra 0} \Xi(\kv)$ of the noise of eq.~\eqref{eq:noise-yeo}.
This form of noise has the advantage of being readily applicable to inhomogeneous systems---however, at the expense of accepting equilibration errors at higher wavenumbers unless simulations are run in a rather small range of parameters, as will be demonstrated in section \ref{sec:results}.
If, on the other hand, only the correct behavior of the model at the largest length scales (i.e.\ the hydrodynamic limit) is of interest, the use of spatially uncorrelated noise is sufficient. Although it fulfills the FDT of fluctuating hydrodynamics strictly only for $k \ra 0$, one can expect it to give still acceptable results almost up to $k\sim 1$.

The fluid momentum in a simulation subject to thermal noise should obey Gaussian statistics in real space with a variance given by $\bra j_\alpha j_\beta\ket = \rho_0 k_B T \delta_{\alpha\beta}$. Hence, as a first check in simulation whether uncorrelated noise is sufficient, one can track the momentum variance computed over the lattice in real-space and compare with the theoretically expected result. Since violated equipartition at smaller scales will inevitably show up in the globally computed equal-time variance, a deviation from the expected value indicates a length-scale dependent dissipation mechanism not captured by spatially uncorrelated noise. Consequently, the use of correlated noise would be required.
More detailed assessments can be performed by computing wavenumber-resolved variances (see section \ref{sec:results}).

Due to the existence of cross-correlations between stress and ghost noises, indicated by a non-zero  $\Xi_{49}$, the noise construction becomes slightly more involved compared to the ideal gas case, where the covariance matrix is diagonal. However, since the dynamics of ghost modes are immaterial for the hydrodynamic limit of a LB model, one might argue that in this limit, one can drop the off-diagonal components of $\Xi$ and proceed using ``ideal gas-like'' noise, with variance defined by the diagonal elements of $\Xi$. Simulations have indeed shown that this is a feasible option, leading to satisfactory equilibration of all LB modes for small wavenumbers.
In this work, however, we shall stick to the exact expression for $\Xi$. In this case, noise variables $\xi_a$ that have the required (non-diagonal) covariance $\Xi(\bv{0})$ can be constructed independently on each lattice site by a standard method \cite{numerical_recipes}, which consists of Cholesky-factorizing the noise matrix as
$\Xi(\b0) = L L^T$, with a left-triangular matrix $L$.
The noise variables follow as $\xi_a=L\nu_a$, where the $\nu_a$ are a set of independent Gaussian random variables of unit variance.
Note that, since $\Xi(\b0)$ is only positive semi-definite, the Cholesky factorization has to be performed effectively with the lower-right $(n-d-1) \times (n-d-1)$ block matrix of $\Xi$. Alternatively, one can compute the Jordan decomposition of $\Xi$, that is $\Xi = S\text{diag}(e_i) S^T$, where $S$ is the orthogonal matrix of the eigenvectors of $\Xi$ and the $e_i$ are the eigenvalues. The required matrix $L$ is then given by $L=S\, \text{diag}(\sqrt{e_i})$.

In cases where uncorrelated noise leads to strong violations of equipartition or where the correct behavior of the model at smaller scales is important, one must use spatially \emph{correlated} noise, i.e.\ implement the exact FDT \eqref{eq:noise-yeo} at each point in $k$-space.
The algorithmic construction of the noise can be performed in Fourier space, as described, for example, in \cite{ojalvo_noise_book}. Here, one Fourier transform back to the real lattice space is required per time-step. The noises $\xi_a(\kv)$ at a point $\kv$ are computed using the same procedure described above for $\kv=\b0$.

\section{Results}
\label{sec:results}

\subsection{Equilibration tests}
As a first benchmark test, we check whether the noise defined by eq.~\eqref{eq:noise-yeo} leads to the correct equilibration of all the LB modes in a simulation.
For this purpose, we perform simulations in a two-dimensional, periodic, \emph{homogeneous} one-phase system of size $128\times 128$ l.u., using either the full (spatially correlated) noise or its $k=0$ (uncorrelated) approximation.
All parameters are chosen such that numerical stability is ensured also if a two-phase interface were present. This imposes, in particular, a lower bound of approximately 4 l.u.\ on the interface width, eq.~\eqref{eq:interf-width}. We use a fluctuation temperature $T=10^{-7}$ (setting $k_B=1$) for all simulations, which lies well within the intrinsic stability constraint \eqref{eq:low-mach} of LB.
The magnitude of the resulting velocity and density fluctuations is then of the order of $10^{-3}$.
All relaxation times are set to a value of $\tau = 1.0$.
For a linearized, homogeneous non-ideal fluid model, the essential input parameters are the square-gradient parameter $\kappa$ and the thermodynamic speed of sound $c_s$ (which is equal to the generalized speed of sound in the limit $k \ra 0$).
The actual values of the parameters $\rho_L$, $\rho_V$ and $\beta$ used in the bulk free energy density \eqref{eq:f0-bulk} are immaterial in the linearized case, but are also stated for reference.

Results are analyzed by comparing the equal-time correlations $\bra |\delta m_a(\kv)|^2\ket$ of a LB mode to the theoretical expectation as expressed by the correlation matrix $G$, eq.~\eqref{eq:g-mat-yeo}. This comparison is most conveniently performed by computing the equilibration ratio, which is defined as $\bra |\delta m_a(\kv)|^2\ket\st{sim} / G_{aa}(\kv)$. In the plots below, this quantity is shown as an average over 400 simulation snapshots.

\begin{figure*}[ht]\centering
    (a)\includegraphics[width=0.32\linewidth]{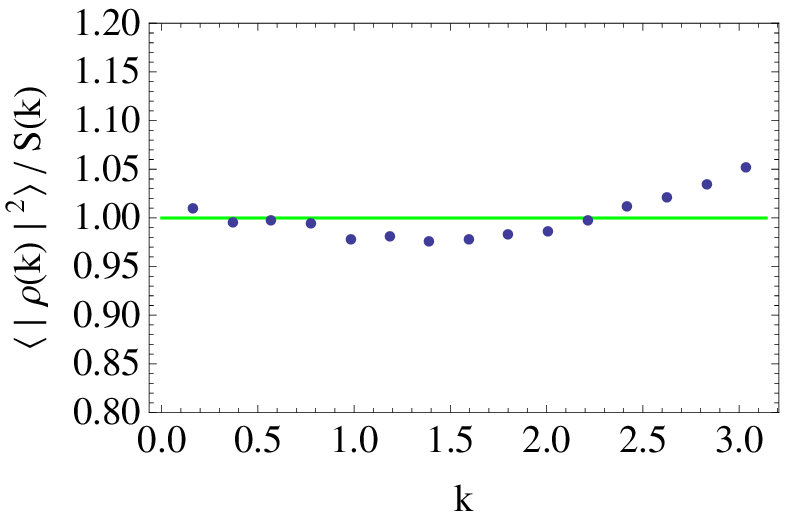}\quad
    (b)\includegraphics[width=0.25\linewidth]{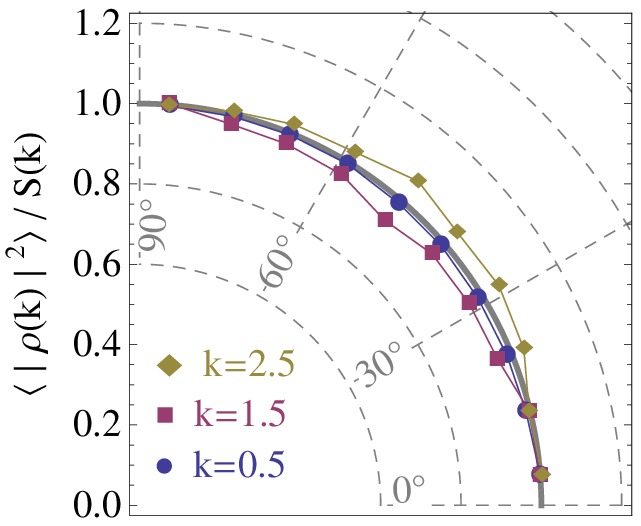}\quad
    (c)\includegraphics[width=0.32\linewidth]{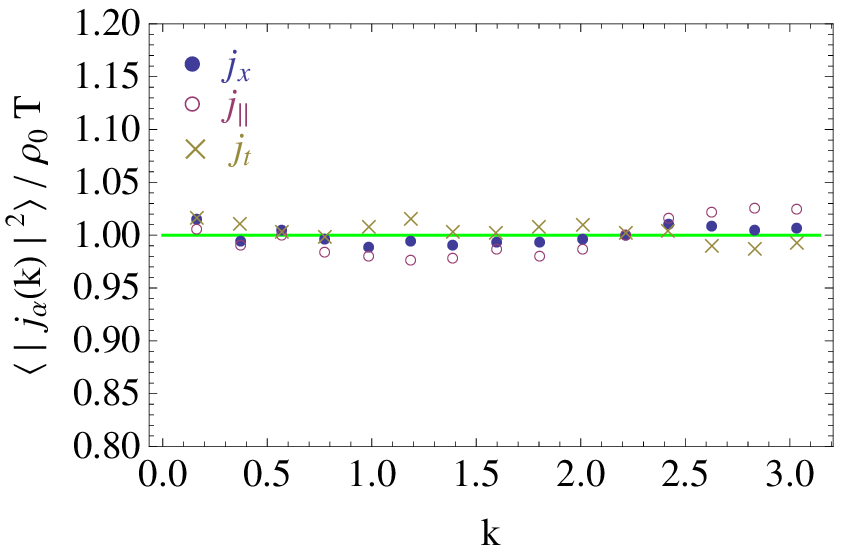}\quad (d)\includegraphics[width=0.25\linewidth]{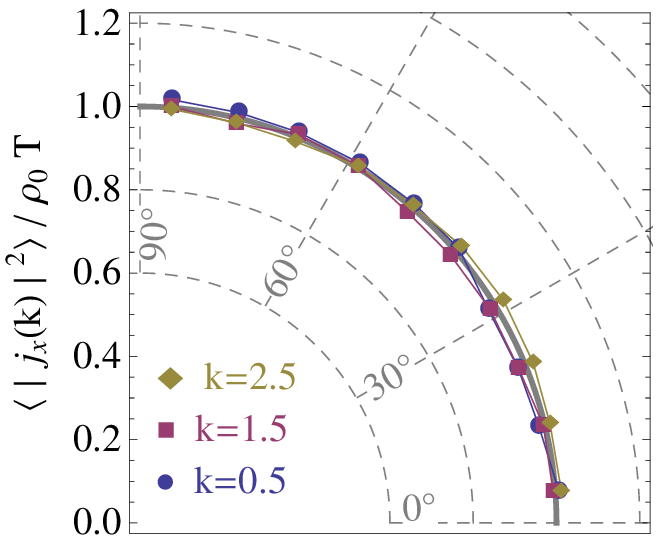}\quad
    (e)\includegraphics[width=0.32\linewidth]{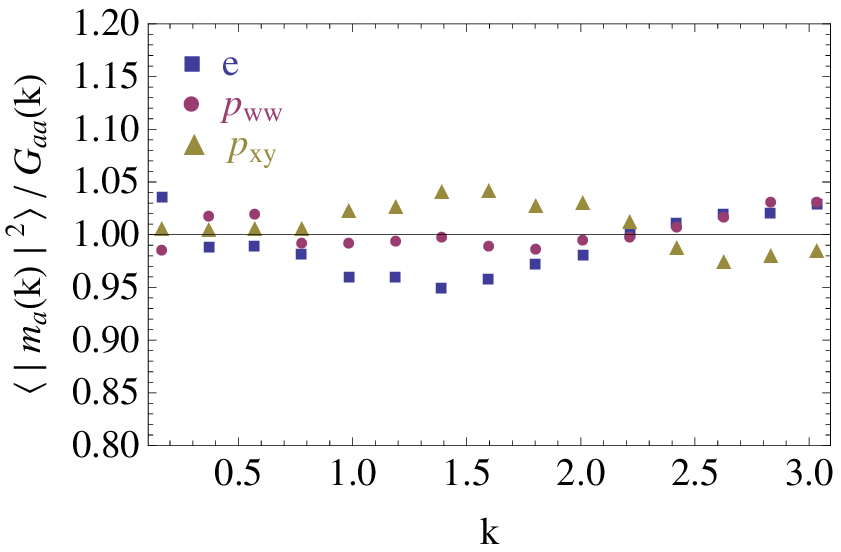}\quad
    (f)\includegraphics[width=0.32\linewidth]{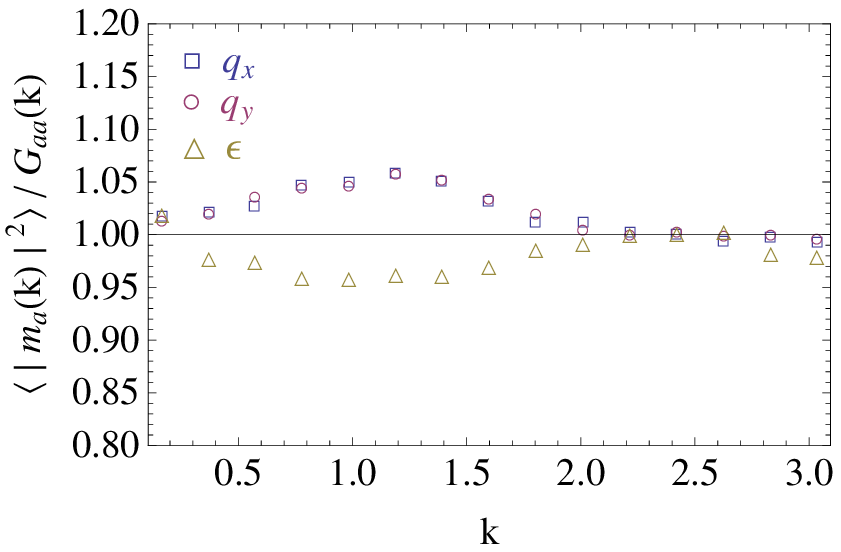}
   \caption{(Color online) Correlated noise in the modified-equilibrium model for $\kappa=0.08$, $c_s=0.265$, $\tau=1.0$. Equilibration ratios of (a,b) the density, (c,d) the momentum, (e) the transport and (f) the ghost modes. In (b) and (d) the dependence of the equilibration ratio of the density and momentum on $\theta$, when $\kv=(\cos\theta,\sin\theta)k$, is shown for several magnitudes of $k$. In the remaining plots, each data point represents an average over all directions in $k$-space at each magnitude $|k|$. $j_x$ denotes the $x$-component, $j_{||}$ the longitudinal and $j_t$ the transverse component (with respect to $\kv$) of the momentum $\bv{j}$.}
    \label{fig:correl-fluct-kappa008}
\end{figure*}
In Fig.~\ref{fig:correl-fluct-kappa008}, the equilibration ratios obtained with the exact (spatially \emph{correlated}) form of noise are shown for $\kappa=0.08$ and $c_s=0.27$ (with the corresponding parameters in the bulk free energy being $\rho_V = 0.5$, $\rho_L=1.0$, $\beta=0.14$).
We see that the maximum error in the equilibration of every mode stays always below 5\%, even for the largest wavenumbers. Quantitatively very similar results have been obtained for all tested combinations of $\kappa$, $c_s$ and $\tau$. In all cases, the equilibration error at intermediate and large wavenumbers is found to remain less than 10\% (occasionally 20\%), while it is generally negligible for smaller wavenumbers where all modes appear perfectly equilibrated.
These results suggest that the noise covariance matrix \eqref{eq:noise-yeo}, and hence the underlying equilibrium correlation matrix $G$ \eqref{eq:g-mat-yeo}, correctly describe the dissipation and the fluctuations in the modified-equilibrium model.

\begin{figure*}[ht]\centering
    (a)\includegraphics[width=0.32\linewidth]{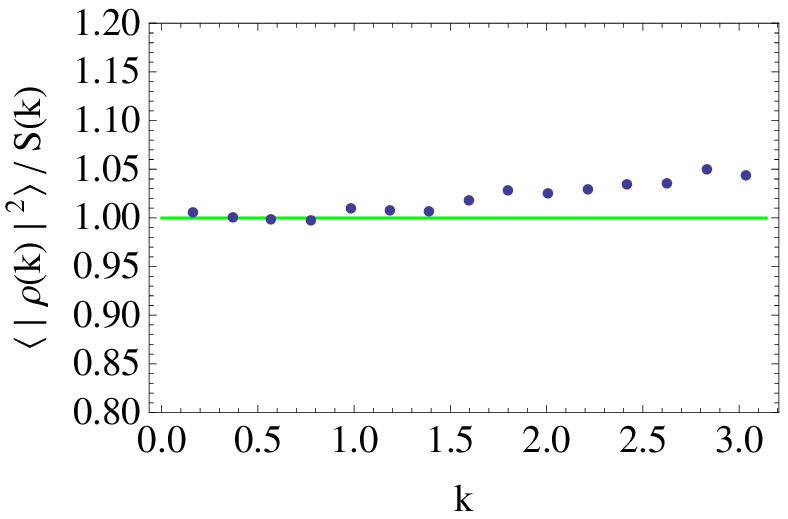}\quad
    (b)\includegraphics[width=0.25\linewidth]{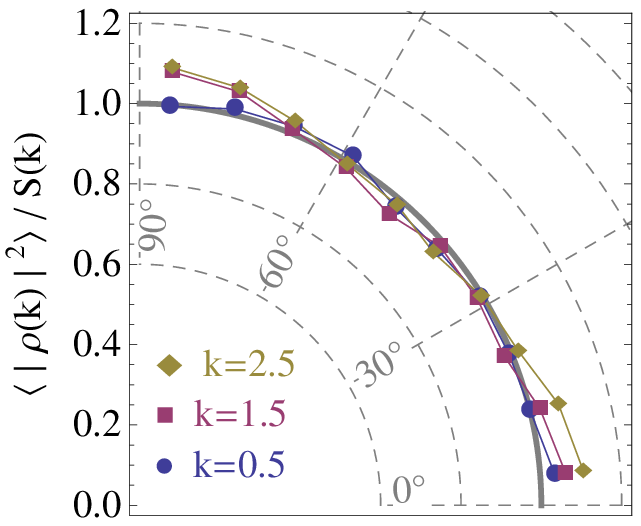}\quad
    (c)\includegraphics[width=0.32\linewidth]{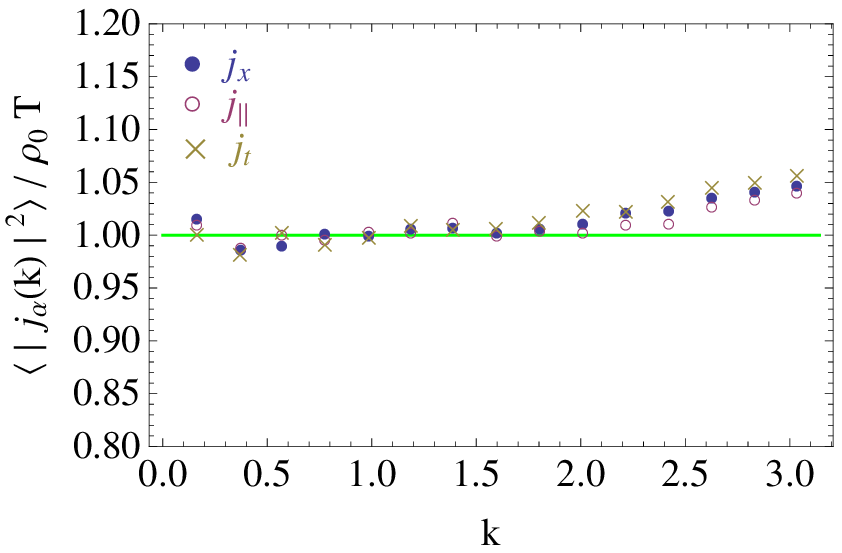}\quad (d)\includegraphics[width=0.25\linewidth]{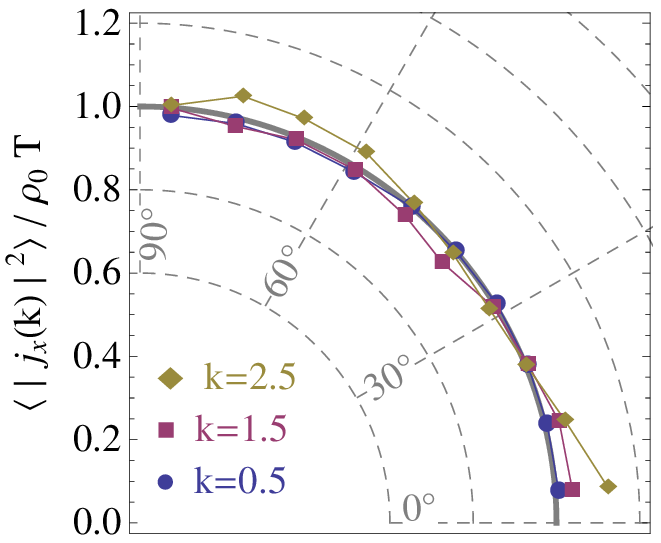}\quad
    (e)\includegraphics[width=0.32\linewidth]{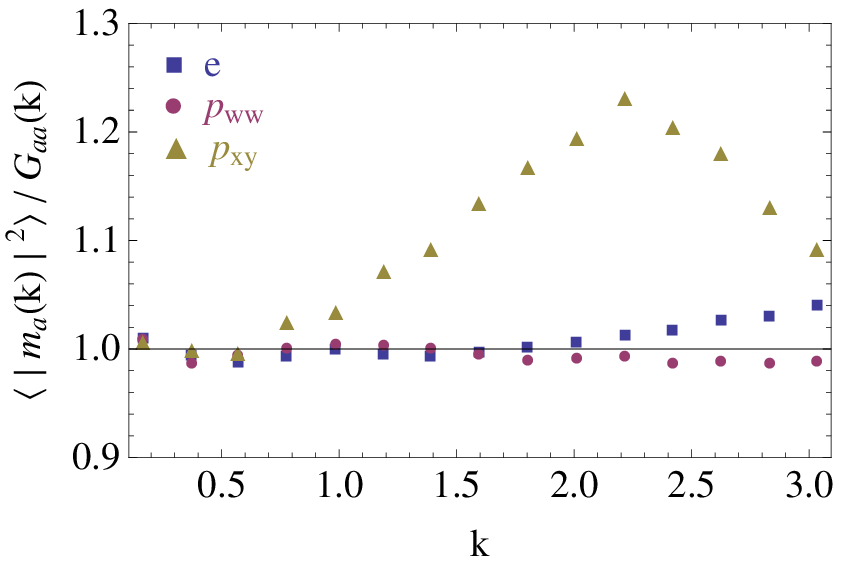}\quad
    (f)\includegraphics[width=0.32\linewidth]{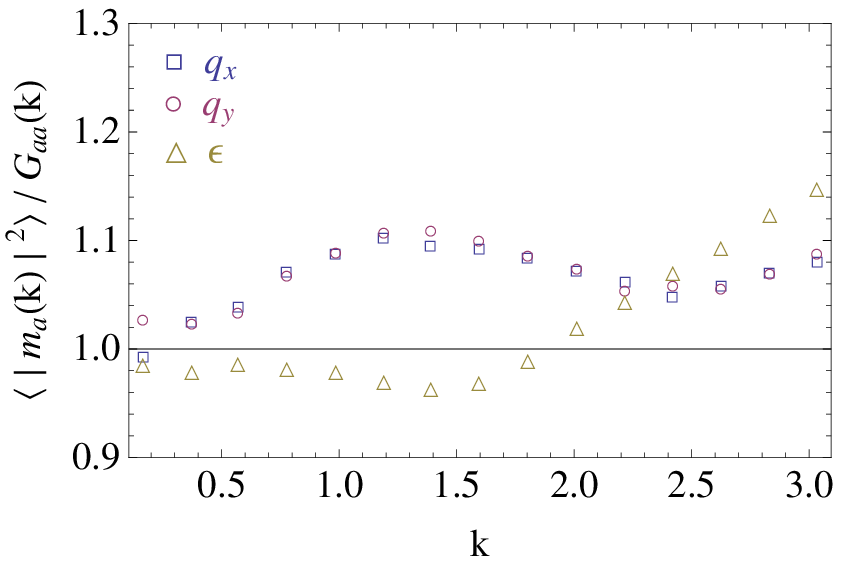}
   \caption{(Color online) Uncorrelated noise in the modified-equilibrium model for $\kappa=0.03$, $c_s=0.15$, $\tau=1.0$. Equilibration ratios of (a,b) the density, (c,d) the momentum, (e) the transport and (f) the ghost modes. In (b) and (d) the dependence of the equilibration ratio of the density and momentum on $\theta$, when $\kv=(\cos\theta,\sin\theta)k$, is shown for several magnitudes of $k$. In the remaining plots, each data point represents an average over all directions in $k$-space at each magnitude $|k|$. $j_x$ denotes the $x$-component, $j_{||}$ the longitudinal and $j_t$ the transverse component (with respect to $\kv$) of the momentum $\bv{j}$.}
    \label{fig:uncorrel-fluct-kappa003}
\end{figure*}
In contrast, using spatially \emph{uncorrelated} noise with the above choice of simulation parameters has been found to lead to large errors at higher wavenumbers.
However, at least for certain choices of $\kappa$ and $c_s$, satisfactory results at all wavenumbers \emph{can} be achieved also with uncorrelated noise.
This is demonstrated in Fig.~\ref{fig:uncorrel-fluct-kappa003}, where equilibration ratios obtained for $\kappa=0.03$ and $c_s=0.15$ are shown (with the corresponding parameters in the bulk free energy \eqref{eq:f0-bulk} chosen as $\rho_V = 0.1$, $\rho_L=1.0$, $\beta=0.015$).
We see that the maximum error in the equilibration of the density and momentum modes is always below 10\%, even for the largest wavenumbers.
The errors in the stress and ghost modes, however, are significantly larger, especially for intermediate $k$. These deviations are found to diminish if spatially correlated noise is used.
In the hydrodynamic region, which is located between $k=0$ and $\sim$ 0.8 for the present choice of parameters \cite{gross_hydro_2010}, all errors can be considered as negligible. This is found to be true for all tested parameter combinations, and holds even for those cases where no acceptable thermalization at larger wavenumbers can be obtained with uncorrelated noise.

\subsection{Capillary fluctuations}
In the presence of interfaces, the fluctuations in the bulk fluid can induce capillary (or interfacial height) fluctuations \cite{grant_desai_1983}.
The static spectrum of large-wavelength, small-amplitude fluctuations of the interface height $h$ is given by \cite{grant_desai_1983, RowlinsonWidom_book}
\beq \bra |h(\kv)|^2 \ket = \frac{k_B T}{\sigma k^2}\,, \label{eq:capill-static}\eeq
where $\sigma$ is the surface tension, eq.~\eqref{eq:surf-tension}.
In order to test whether this relation can be reproduced by the fluctuating non-ideal fluid model, we perform simulations of a liquid-vapor interface that belongs to an extended liquid stripe placed in a fully periodic, two-dimensional box of size $L_x\times L_y= 2048\times 400$ l.u. The liquid stripe has a width of $180$ l.u.\ and is aligned parallel to the $x$-axis. Simulation parameters are $\rho_L=1.0$, $\rho_V=0.5$, $\beta=0.04$, $\kappa=0.03$, $T=10^{-7}$ and $\tau=0.1$.
For the simulation of capillary fluctuations with the modified-equilibrium model, Galilean-invariance correction terms become important; we use the expressions proposed by \cite{holdych_gal_1998}.
The interface height $h(x)$ is defined as the position of the point where the density equals $(\rho_L+\rho_V)/2$.
Spatially uncorrelated Langevin noise defined by $\Xi(\b0)$, eq.~\eqref{eq:noise-yeo}, is employed. In order to apply this noise to inhomogeneous situations, such as the present one, the local values of the density $\rho_0$ and the speed of sound $c_s$ have to be taken into account in computation of the real-space noise variance.

\begin{figure}[hb]\centering
    \includegraphics[width=0.8\linewidth]{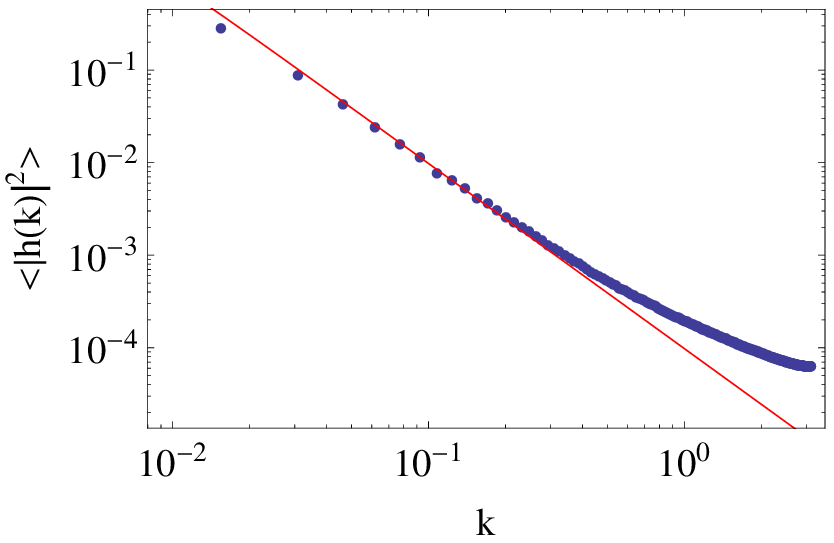}
   \caption{(Color online) Capillary fluctuations of a planar interface in the modified-equilibrium model using spatially uncorrelated noise. The equal-time spectrum of interfacial height fluctuations obtained from simulation [dots] is compared to the theoretical capillary structure factor [solid line, eq.~\eqref{eq:capill-static}]. $k$ denotes the wavenumber along the interface. Simulation parameters: $\kappa=0.03$, $\beta=0.04$, $\rho_L=1.0$, $\rho_V=0.5$, $\tau=0.1$, the interface width is approximately 5 l.u.}
    \label{fig:capill-static}
\end{figure}

Fig.~\ref{fig:capill-static} shows the spectrum of the interfacial height fluctuations obtained for the modified-equilibrium model. We see, that the simulation results are well described by the theoretical capillary structure factor \eqref{eq:capill-static} for wavenumbers $k\lesssim 0.2$
\footnote{Simulations using ideal gas-type noise for the force-based model of Lee and Fischer \cite{lee_fischer_2006} gave similar results, agreeing with the capillary structure factor up to a slightly larger value of $k\sim 0.5$.}.
We ascribe the deviations at larger wavenumbers to the presence of error terms in the hydrodynamic equations of the model and to the fact that the harmonic approximation, on which expression \eqref{eq:capill-static} is based, breaks down and curvature corrections as well as lattice effects become important.
At the smallest wavenumbers, fluctuations are damped due to compressibility of the liquid stripe as a whole.

\section{Summary and Outlook}
We have presented in this paper a systematic study of thermal fluctuations in the LB method for non-ideal fluids.
As a specific example, we considered the Langevin version of the modified-equilibrium model proposed by Swift et al., which approximates the stochastic Navier-Stokes equations for non-ideal fluids based on square-gradient free energy functionals.
Thermal fluctuations are implemented by adding Gaussian random noise sources to the deterministic LBE, thereby promoting it to a LB-Langevin equation.
In order to determine the covariance matrix of the noise, a fluctuation-dissipation theorem has been derived, which requires as input an expression for the equal-time correlations of the distribution function in a non-ideal LB fluid.
Drawing on continuum kinetic theory of fluids, a general ansatz [eq.~\eqref{eq:f-correl-lb}] for these correlations is provided.
Application of this ansatz to the modified-equilibrium LB model requires non-trivial modifications to ensure that the noise covariance matrix is positive-semidefinite and obeys the FDT of fluctuating hydrodynamics at finite wavenumbers.
We have obtained a wavenumber-dependent form of thermal noise that was shown to lead to excellent thermalization of all modes in a simulation at all wavenumbers and all tested parameters.
The necessity of spatially correlated noise for the modified-equilibrium model, although a priori unexpected, can be traced back to the fact that this model is based on a non-Maxwellian form of the equilibrium distribution function, in contrast to standard kinetic theory.

Additionally, we have demonstrated that thermal fluctuations in the hydrodynamic regime can often already be satisfactory modeled using noise evaluated in the zero wavenumber-limit.
This result is expected to hold strictly whenever non-ideal fluid interactions enter the hydrodynamic equations reversibly, in which case the random stress tensor is independent of wavenumber. The modified-equilibrium model only approximately fulfills this requirement, as the non-ideal interactions---even though weakly---do contribute to the dissipative terms in this model.
Since noise obtained in the zero wavenumber-limit is spatially uncorrelated by construction, it has the advantage of being easy to implement and readily applicable to inhomogeneous systems.
In the small wavenumber-regime, good equilibration of all LB modes has been obtained for all simulation parameters.

For a future study, it would be interesting to apply the presented approach also to force-based non-ideal fluid models. Preliminary results using the model of Lee and Fischer \cite{lee_fischer_2006} indicate that acceptable equilibration for low wavenumbers can be obtained using noise identical to the one employed for the ideal gas \cite{gross_hydro_2010}. This finding agrees with the expectations based on the corresponding FDT in the hydrodynamic limit.

Although we assumed that density fluctuations are described by a square-gradient free energy functional, the present theory puts in fact no constraints on the form of the structure factor. Hence, our approach should in principle also be applicable to models based on different thermodynamic approaches, such as \cite{shan_lattice_1993}.
Along this direction, the extension of the present theory to multi-component approaches such as, for example, binary fluids \cite{swift_lattice_1996}, or emulsions \cite{benzi_glassyLB_2009} might be particularly interesting.
Finally, it might be interesting to investigate how the noise for a non-ideal LB fluid can be derived without resorting to results of continuum kinetic theory, but instead using the approach to the fluctuating LBE proposed by D\"unweg et al.\ \cite{duenweg_statmechLB_2007}.

\section{Acknowledgments}
We thank Kevin Stratford, P.T.\ Sumesh and Alexander Wagner for useful discussions. M.G.\ thanks the EPCC for hospitality during the inception of this work and acknowledges financial support from HPC-Europa2 Project (Project No.\ 228398). M.E.C. acknowledges support from the Royal Society and funding from EPSRC Grant No.\ EP/E030173. M.G.\ and F.V.\ further acknowledge financial support by the Deutsche Forschungsgemeinschaft (DFG) under the Grant No.\ Va205/3-3 (within the Priority Program SPP1164) as well as funding from the industrial sponsors of ICAMS, the state of North-Rhine Westphalia and the European Commission in the framework of the European Regional Development Fund (ERDF).

%----------------------------------------------------------------------------------------
\appendix

\section{Kinetic theory of non-ideal fluids}
\label{sec:f-correl}
We present here a brief account on the kinetic theory of fluctuations in a non-ideal fluid and show how this theory can be applied to the LB method and the particular model considered in section \ref{sec:yeo-model}.

\subsection{Continuum theory}
In a kinetic description of a non-ideal fluid, the knowledge of the density and momentum correlations [eqs.~\eqref{eq:struct-fact} and \eqref{eq:momtm-fluct}] alone is not sufficient. Instead, one must specify also the equal-time correlations of the fluctuations in the one-particle distribution function $f(\rv,\cv)$, where $\cv$ is the molecular velocity. The corresponding expression can be motivated from statistical mechanics within Klimontovich's approach to kinetic theory \cite{klimontovich_nonideal_1973, liboff_book}.

This approach is based on defining a phase-space density as
\beqn
F(\rv,\cv)=\mu \sum_i \delta(\rv-\rv_i)\delta(\cv-\cv_i)\,,
\eeqn
where $\mu$, $\rv$ and $\cv$ refer to the mass, positions and velocities of the fluid particles, respectively.
Since in our case the relevant quantity is the mass density instead of the number density, we have defined the phase density with an additional factor of mass. This ensures that the one-particle distribution functions derived below are defined in terms of mass density, in agreement with the situation in LB.
Reduced particle distribution functions $f^n$ can be defined by computing moments of $F$ with respect to the full $N$-particle distribution function $f^N$. The first two reduced distribution functions are given by \cite{liboff_book}:
\begin{align}
\bra F(\rv,\cv) \ket &= f_1(\rv,\cv)\\
\begin{split}
\bra F(\rv,\cv)F(\rv',\cv') \ket &= f_2(\rv,\cv,\rv',\cv') \\ &\quad + \mu \delta(\rv-\rv')\delta(\cv-\cv') f_1(\rv,\cv)
\end{split}
\label{eq:FF-correl}
\end{align}

We consider now fluctuations $\delta F = F - \bra F\ket$ around a global equilibrium state with density $\rho_0$ and zero flow velocity, i.e. $\bra F\ket = \bar f_1(\cv)$ is assumed to be a global Maxwellian distribution.
The fluctuation $\delta F$ can then be interpreted as a fluctuation $\delta f_1$ in the one-particle distribution function $f_1$ over a uniform reference state described by $\bar f_1$.
In a Langevin description, we thus promote $f_1$ to be an instantaneously fluctuating quantity given by $f_1(\rv,\cv) = \bar f_1(\cv) + \delta f_1(\rv, \cv)$.
Specializing to a translationally invariant system, eq.~\eqref{eq:FF-correl} allows to determine the equal-time correlations of the fluctuations in the one-particle distribution function as
\beq\begin{split}
\bra \delta f_1(\rv,\cv) \delta f_1(\rv',\cv')\ket &= \bra \delta F(\rv,\cv) \delta F(\rv',\cv')\ket \\
&= \bra F(\rv,\cv) F(\rv',\cv')\ket -  \bar f_1(\cv) \bar f_1(\cv')\\
&= \bar f_1(\cv) \bar f_1(\cv') [g(\rv-\rv')-1] \\&\quad + \mu \delta(\rv-\rv')\delta(\cv-\cv') \bar f_1(\cv)\,,
\label{eq:f1-fluct-real}
\end{split}\eeq
where in the last step, we introduced the pair correlation function $g$ by $f_2(\rv-\rv',\cv,\cv') = f_1(\cv) f_1(\cv') g(\rv-\rv')$.
The structure factor, defined in terms of the relative fluctuations, $S(\rv) = \bra \delta \rho(\rv)\delta \rho \ket$ [cf.\ \eqref{eq:struct-fact}], is related to the pair correlation function by \cite{Chaikin_book, Hansen_ThoSL}
\beqn S(\rv) = \rho_0^2 (g(\rv)-1) + \mu \rho_0 \delta(\rv)\,.
\eeqn
Transforming relation \eqref{eq:f1-fluct-real} to Fourier space and dropping the index 1,
we finally obtain the desired relation
\begin{multline}
\bra \delta f(\kv,\cv) \delta f(\kv',\cv')\ket =  \Big[ \bar f(\cv) \bar f(\cv') [S(\kv)/\rho_0-\mu]/\rho_0 \\ + \mu \bar f(\cv) \delta(\cv-\cv') \Big] \delta(\kv+\kv')\,.
\label{eq:f-correl-cont}
\end{multline}
The first term on the right hand side of \eqref{eq:f-correl-cont} describes spatial correlations due to the non-ideal character of the fluid. For the ideal gas, $S(\kv)=\mu \rho_0$, thus only the last term remains, and relation \eqref{eq:f-correl-cont} becomes identical to the expression used in the Boltzmann-Langevin theory for a dilute gas \cite{bixon_zwanzig_1969, fox_uhlenbeck_1970b, Landau_PhysKin}.
As can be easily checked, relation \eqref{eq:f-correl-cont} contains already expression \eqref{eq:struct-fact} for the structure factor and expression \eqref{eq:momtm-fluct} for the momentum correlations, since, by definition, $\rho = \int f(\cv) d^d \cv$ and $\bv{j} = \int f(\cv) \cv\; d^d\cv$.

\subsection{LB theory}
The LB analog of eq.~\eqref{eq:f-correl-cont} for the equal-time correlations of fluctuations in the distribution function of a non-ideal fluid, can be written as [eq.~\eqref{eq:f-correl-lb}]
\beq
\bra \delta f_i(\kv) \delta f_j(\kv') \ket = \Big[\bar f_i \bar f_j [S(\kv)/\rho_0-\mu]/\rho_0 + \mu \bar f_i \delta_{ij}\Big] V\delta_{\kv,-\kv'}\,,
\label{eq2:f-correl-lb}
\eeq
where $\bar f_i = f\ueq_i(\rho_0, u=0)$ is the distribution function of a quiescent reference state. In this context, the parameter $\mu$ can be interpreted as the mass of a fictitious fluid particle. Its value is a priori unknown and thus has to be determined such that the correct momentum variance as required by statistical mechanics, eq.~\eqref{eq:momtm-fluct}, is obtained. The factor $V$, representing the system volume, arises by dimensional consistency and will be neglected henceforth.
The correlation matrix $G$ is obtained from \eqref{eq2:f-correl-lb} through a basis transformation,
\beq \begin{split}
G_{ab}(\kv) &= T_{ai} T_{bj} \bra \delta f_i(\kv) \delta f_j(-\kv) \ket \\
&= \bar m_a \bar m_b [S(\kv)/\rho_0-\mu]/\rho_0 + \mu T_{ai} T_{bi} \bar f_i \,.
\label{eq2:g-correl}
\end{split}\eeq

We shall briefly demonstrate how the unknown parameter $\mu$ can be determined for the example of an ordinary \emph{ideal gas} LB model. In this case, we have $\bar f_i = \rho_0 w_i$, $\sum_i \cv_i \bar f_i = 0$ and hence
\beqn
\begin{split}
\bra j_{\alpha} j_{\beta} \ket &= c_{i\alpha} c_{k\beta} \bra \delta f_i \delta f_k \ket =  c_{i\alpha} c_{i\beta} \mu \bar f_i = \mu\rho_0 \cslb^2 \delta_{\alpha\beta} \,,
\end{split}\eeqn
where additionally the orthogonality of the basis vectors and relation \eqref{eq:scalar-prod} have been invoked.
In order to obtain the desired expression $\bra j_\alpha j_\beta\ket = \rho_0 k_B T \delta_{\alpha\beta}$, we see that the mass parameter must be chosen as $\mu = k_B T / \cslb^2$, in agreement with the ideal gas equation of state.
An analogous calculation of the density correlator shows that $\mu$ is in fact the structure factor of the ideal LB gas divided by $\rho_0$, $\mu = S\st{id,LB}/\rho_0$.

A strict application of eq.~\eqref{eq2:g-correl} to compute the equilibrium correlation matrix $G$ for the \emph{modified-equilibrium model} requires to use the equilibrium moments (Table~\ref{tab:mom-yeo}) evaluated in a quiescent state, $\bar m_a = m_a\ueq(\rho_0, \mbox{u=0}) = \{\rho_0, 0,0, d(\b0)\rho_0, 0,0,0,0, -d(\b0)\rho_0\}$, where $d(\b0)\equiv \lim_{k\ra 0} d(\kv)$.
The requirement that $G_{11}=S(\kv)$ and $G_{22}=G_{33}=\rho_0 k_B T$ fixes
$\mu= k_B T /c_s^2 \,.$
However, it can be shown that if the equilibrium correlation matrix $G$ constructed in this way is used, the noise following from eq.~\eqref{eq:noise-matrix} \emph{violates} the FDT fluctuating hydrodynamics at any \emph{finite} wavenumber.
This can be understood from the fact that the bulk viscosity of the modified-equilibrium model is given by eq.~\eqref{eq:bulkv-yeo}, $\zeta(\kv) = \rho_0 \cslb^2 \left(\tau_b-\onehalf\right)\left[2-\frac{c_s^2(\kv)}{\cslb^2}\right],$
and---in particular---is dependent on wavenumber.
As shown in appendix \ref{sec:fdt-hydro-lim}, the FDT of fluctuating hydrodynamics requires the same factor $2-c_s^2(\kv)/\cslb^2$ to appear in variance of the LB noise pertaining to the bulk stress mode, $\bra \xi_4^2 \ket$.
The above $G$, however, would lead to an expression for $\bra \xi_4^2\ket$ that is correct only at $k=0$.
Moreover, a closer investigation of the so obtained noise matrix reveals the presence of a negative eigenvalue for finite wavenumbers---a fact that invalidates its meaning as a covariance matrix and inhibits its use in simulation.

All the above mentioned problems can, however, be successfully solved by replacing all occurrences of $c_s^2$ in $G(\kv)$ by the full $k$-dependent speed of sound $c_s^2(\kv) = \rho_0 k_B T/S(\kv)$. This is tantamount to re-introducing the $k$-dependent terms $d(\kv)$ in $\bar m_a$, and then using the so defined $k$-dependent reference state $\bar m_a(\kv)$ in the computation of $G$ according to \eqref{eq2:g-correl}. Agreement with the statistical mechanical expression for the momentum correlation function now requires to introduce a $k$-dependent ``mass'' parameter $\mu$ as
\beqn \mu(\kv) = k_B T/c_s^2(\kv) = S(\kv)/\rho_0\,.\eeqn
Inserting this expression for $\mu(\kv)$ into \eqref{eq2:f-correl-lb} then finally leads to relation \eqref{eq:f-correl-yeo}.

The above derivation shows that, while equation \eqref{eq2:f-correl-lb} is a useful starting point to arrive at a valid expression for the correlations of the distribution function, it may have to be modified when applied to a specific non-ideal LB fluid. In the case of the model of Swift et al.\ we considered above, the modifications can be traced back to the fact that the local equilibrium distribution is defined in a non-standard way compared to continuum kinetic theory, where one assumes a Maxwellian form. Hence, we expect relation \eqref{eq2:f-correl-lb} in its original form to be more appropriate to non-ideal fluid LB models that employ the usual ideal gas equilibrium distribution. This will be investigated in future works.

\section{Lattice Fourier transforms}
\label{sec:fourier-latt}
Working with Fourier transformations on a lattice requires to take into account the correct equivalents of the discretized derivative operators. In the present case, a discretized Laplace operator of the form
\beqn
\sum_{i\neq 1} w_i \left[ \rho(\bv{r} + \bv{c}_i) + \rho(\bv{r}-\bv{c}_i) - 2\rho(\bv{r}) \right]/\cslb^2\,
\eeqn
is used,
where $i$ runs over all eight non-zero directions of the D2Q9 lattice.
The Fourier transformed discretized Laplace operator follows as
\begin{multline*}
\left(\frac{4}{9}(\cos k_x + \cos k_y) + \frac{2}{9}(\cos k_x \cos k_y) -\frac{10}{9}\right)/\cslb^2\,.
\end{multline*}
This expression has to be used in place of $-k^2$ in the LB analogs of the structure factor and the speed of sound.
Note that, in order not to clutter-up notation, we prefer to state the continuum expressions throughout the main text.
In general, deviations between the continuum and discrete derivative operators become significant only at intermediate and high wavenumbers. There, isotropy is typically lost, i.e., the lattice Laplace operator depends on the direction in $k$-space.

\section{Fluctuating stress tensor}
\label{sec:fdt-hydro-lim}
In the hydrodynamic limit, it is possible to directly compute the noise strength of the stress modes required by the FDT of fluctuating hydrodynamics, eqs.~\eqref{eq:ns-random-stress-l} and \eqref{eq:ns-random-stress-l}. In Fourier space, the random stress tensor correlations can be expressed as \cite{Reichl_book}
%\begin{widetext}
%\beq
\begin{multline}\bra R_{\alpha \beta}(\kv,t) R_{\gamma \delta}(\kv',t') \ket = 2k_B T \Big[ \eta \Big(\delta_{\alpha \gamma} \delta_{\beta \delta} + \delta_{\alpha \delta} \delta_{\beta \gamma} \\- \frac{2}{d} \delta_{\alpha \beta}\delta_{\gamma \delta}\Big)
+ \zeta\, \delta_{\alpha \beta}\delta_{\gamma \delta}\Big]\delta_{\kv,-\kv'}\delta_{t,t'},
\label{eq:fluct-hydro-fdt-latt}
\end{multline}%\eeq
%\end{widetext}
where the viscosities are in principle allowed to depend on the magnitude of the wavenumber $k$.
On the other hand, a Chapman-Enskog analysis performed on the fluctuating LBE \eqref{eq:fluct-lin-lbe} reveals that, for the presently chosen D2Q9 basis set (see Table~\ref{tab:modes-d2q9}), the random stress tensor is related to the noise variables by (cf.\ \cite{duenweg_statmechLB_2007})
\beq
R_{\alpha\beta} = \cslb^2\begin{pmatrix} \frac{\xi_4}{\lambda_b \sqrt{N_4}} + \frac{\xi_5}{\lambda_s\sqrt{N_5}} & \frac{\xi_6}{\lambda_s \sqrt{N_6}} \\
\frac{\xi_6}{\lambda_s \sqrt{N_6}} & \frac{\xi_4}{\lambda_b \sqrt{N_4}} - \frac{\xi_5}{\lambda_s \sqrt{N_5}} \end{pmatrix}\,.
\label{eq:lb-fluct-stress}
\eeq
This result is independent of the particular non-ideal fluid model and only depends on the underlying lattice.
After rearranging \eqref{eq:lb-fluct-stress} to obtain the $\xi_a$ in terms of $R_{\alpha\beta}$, we use \eqref{eq:fluct-hydro-fdt-latt} to evaluate the variances $\bra \xi_a^2\ket$ and finally plug in the known expressions of the shear and bulk viscosity.
We find
\beq \begin{split}
\xi_4 &= 3\lambda_b\tr R\\
\Rightarrow \bra \xi_4^2 \ket &= 9\lambda_b^2 \cdot 8 k_B T \zeta \\&= - 36\, k_B T \rho \cslb^2 (2\lambda_b+\lambda_b^2)h\,,
\label{eq:e-fluct}
\end{split}\eeq
where $h=1$ for an ideal gas-like model and \mbox{$h=2-\csph^2(k)/\cslb^2$} for the modified equilibrium model due to the modified bulk viscosity \eqref{eq:bulkv-yeo},
\beqn \begin{split}
\xi_5 &= \lambda_{s} (R_{xx}-R_{yy})\\
\Rightarrow \bra \xi_5^2\ket &=  \lambda_{s}^2\cdot 8 k_B T  \eta \\
&= -4 k_B T \rho \cslb^2(2\lambda_s + \lambda_s^2)\,,
\end{split}
\eeqn

\beqn \begin{split}
\xi_6 &=  \lambda_s (R_{xy}+R_{yx})/2\\
\Rightarrow \bra \xi_6^2 \ket &=  \lambda_s^2 \cdot 2k_B T  \eta\\
&=-k_B T \rho \cslb^2 (2\lambda_s+\lambda_s^2)\,.
\end {split}\eeqn

% \bibliographystyle{apsrev}
% \bibliography{bibliography}

\end{document}